\documentclass[
    a4paper,
    notitlepage,
    prb,
    twocolumn,
    superscriptaddress,
    floatfix
]{revtex4-1}
\usepackage{amsmath,amssymb}
\usepackage{lipsum}
\usepackage{bm}
\usepackage{graphicx}
\usepackage{graphics}
\usepackage{amsmath}
\usepackage{array}
\usepackage[caption=false]{subfig}
\usepackage{xcolor}
\usepackage{subfig}
\usepackage{hyperref}
\usepackage[capitalise]{cleveref}
\usepackage{textcomp}

\usepackage{epstopdf}
\usepackage{cases}
\usepackage{amssymb}
\usepackage{multirow}
\usepackage{siunitx}
\usepackage{cases}
\usepackage{tabularx}
\usepackage[top=1.1in, bottom=1.1in, left=1.0in, right=1.0in]{geometry}
\usepackage[utf8]{inputenc}
\usepackage{tikz}
\usepackage{comment}
\usepackage{lineno,hyperref}
\modulolinenumbers[5]
\usepackage{booktabs}

\begin{document}

\title{Electron-phonon coupling and non-equilibrium thermal conduction in ultrafast heating systems}
\author{Chuang Zhang}
\email{zhangc520@hdu.edu.cn}
\affiliation{Department of Physics, Hangzhou Dianzi University, Hangzhou 310018, China}
\author{Rulei Guo}
\affiliation{Department of Mechanical Engineering, The University of Tokyo, 7-3-1 Hongo, Bunkyo, Tokyo 113-8656, Japan}
\author{Meng Lian}
\affiliation{School of Physics, Institute for Quantum Science and Engineering and Wuhan National High Magnetic Field Center, Huazhong University of Science and Technology, Wuhan 430074, China}
\author{Junichiro Shiomi}
\email{Corresponding author: shiomi@photon.t.u-tokyo.ac.jp}
\affiliation{Department of Mechanical Engineering, The University of Tokyo, 7-3-1 Hongo, Bunkyo, Tokyo 113-8656, Japan}
\affiliation{Institute of Engineering Innovation, The University of Tokyo, 7-3-1 Hongo, Bunkyo, Tokyo 113-8656, Japan}
\date{\today}

\begin{abstract}

The electron-phonon coupling in ultrafast heating systems is studied within the framework of Boltzmann transport equation (BTE) with coupled electron and phonon transport.
A discrete unified gas kinetic scheme is developed to solve the BTE, in which the electron/phonon advection, scattering and electron-phonon interactions are coupled together within one time step by solving the BTE again at the cell interface.
Numerical results show that the present scheme can correctly predict the electron-phonon coupling constant, and is in excellent agreement with typical two-temperature model (TTM) and experimental results in existing literatures and our performed time-domain thermoreflectance technique.
It can also capture the ballistic or thermal wave effects when the characteristic length/time is comparable to or smaller than the mean free path/relaxation time where the TTM fails.
Finally, the electron-phonon coupling in transient thermal grating geometry and Au/Pt bilayer metals with interfacial thermal resistance is simulated and discussed.
For the former, heat flow from phonon to electron is predicted in both the ballistic and diffusive regimes.
For the latter, the reflected signal increases in the early tens of picoseconds and then decreases with time after the heat source is removed.

\end{abstract}

\maketitle

\section{INTRODUCTION}

Ultrafast laser heating plays a more and more important role in industrial manufacturing of micro/nano electronic devices, medical detection and the exploration of frontier basic science, which involves the interactions between various energy-carrying (quasi)particles~\cite{ChenG05Oxford,kittel1996introduction}.
One of the key factors in multiscale energy transfer and conversion is the coupling between electron and phonon (lattice vibration) in solid materials~\cite{PhysRevLett.68.2834,karna_direct_2023,Sciadv_2019_hotelectron,PhysRevLett.59.1460,TTM_EP_AIPX2022}.
The electron-phonon coupling at the microscopic level can explain many macroscopic thermal and electrical transport phenomena, including thermoelectric conversion ~\cite{mao_thermoelectric_2021,MTP_review_nuo_2021}, electrothermal power consumption and transfer in semiconductor chips~\cite{pop2004analytic,pop_energy_2010} and so on.

Over the past decades, many macroscopic phenomenological heat conduction models are developed to describe the electron-phonon coupling~\cite{PhysRevB.74.024301,PhysRevLett.58.1680,PhysRevB.45.5079,TTM_EP_AIPX2022,PhysRevX.6.021003,PhysRevB.98.134309,zhou_electrohydrodynamics_2015}.
One of the most widely used theoretical models is the two-temperature model (TTM)~\cite{PhysRevLett.58.1680,PhysRevB.50.15337,QIU19942789,PhysRevB.77.075133,PhysRevB.65.214303,TTM_EP_AIPX2022} proposed by Anisimov $et~ al.$~\cite{anisimov1974electron}.
In this empirical model, the electron temperature $T_e$ and phonon temperature $T_p$ are introduced individually and their interactions are represented by a single phenomenological electron-phonon coupling constant $G$.
Although it is simple, it is widely used in ultrafast pump-prope experiments for calculating the electron-phonon coupling constant in metals~\cite{PhysRevB.51.11433,PhysRevLett.59.1962,PhysRevLett.53.1837,Liang_JHT2014,QIU1992719,QIU19942799}.
The Fourier's law is used to express the evolution process of electron and phonon in the spatial and temporal spaces so that the TTM is a parabolic two-step heat conduction equation with infinite heat propagation speed~\cite{RevModPhysJoseph89}.
In order to remove this non-physical assumption, a hyperbolic two-temperature model is developed by introducing a delay term of the derivative of heat flow with respect to time for electron or phonon transport~\cite{QIU19942789,MTM_hyperbolic2021}, which is like an extension of Cattaneo equation~\cite{cattaneo1948sulla}.
In addition, the non-thermal lattice model~\cite{PhysRevX.6.021003} or multitemperature model~\cite{PhysRevB.98.134309,MTM_hyperbolic2021} is developed, in which phonons with different modes or branches are not in local thermal equilibrium and many lattice temperatures and electron-phonon coupling constants are introduced to describe the complex physical interactions.

Although above phenomenological heat conduction models made great success in describing ultrafast energy exchange, they are not suitable to capture the highly non-equilibrium situations, for example when the system characteristic length is comparable to or smaller than electron/phonon mean free path where the diffusive transport is broken~\cite{NHTPA09TTM_BTE,TTM_EP_AIPX2022}.
To study the multiscale energy transfer in ultrafast heating systems, the time-dependent Boltzmann transport equation (BTE)~\cite{IJHMT_2006_EPcoupling_BTE,JHTelectron-phonon2009,PhysRevB.103.125412,PhysRevResearch.3.023072,JAP2016EP_review} with coupled electron and phonon thermal transport becomes an optimal compromise between efficiency and accuracy, which can capture the diffusive and ballistic thermal transport simultaneously.
Instead of directly describe the evolution of macroscopic temperatures, the macroscopic fields are obtained by taking the moment of distribution function within the framework of BTE, so that the key of BTE is tracing the evolution of phonon/electron distribution function in seven-dimensional phase space (time, position, momentum).
Two kinds of numerical methods are usually used to solve the BTE with coupled electron and phonon thermal transport.
One is the statistics Monte Carlo method~\cite{RevModPhys.55.645,MUTHUKUNNILJOSEPH2022107742}, and the other is the discrete ordinate method~\cite{ADAMS02fastiterative,PhysRevB.103.125412} which discretizes the whole three-dimensional phase spaces into a lot of small pieces.
Both of them made progress in highly non-equilibrium heat transfer problems.
However the advection and scattering are treated separately in single time step at the numerical discrete level, so that they have large numerical dissipations in the diffusive regime.

In this work, non-equilibrium thermal conduction in ultrafast heating systems is studied by the BTE accounting for electron-phonon coupling, and a discrete unified gas kinetic scheme (DUGKS)~\cite{guo_progress_DUGKS,zhang_discrete_2019} is developed in which the electron/phonon advection, scattering and electron-phonon interactions are coupled together within one time step by solving the BTE again at the cell interface.
Numerical results show that the present results are in excellent agreement with the experiments and TTM results in the diffusive regime, and can capture the ballistic or thermal wave effects.

The remainder of this article is organized as follows.
The BTE for coupled electron and phonon thermal transport and the DUGKS solver are introduced in Sec.~\ref{sec:BTEtheory} and Sec.~\ref{sec:DUGKS}, respectively.
Results and discussions are shown in Sec.~\ref{sec:results}.
Finally, a conclusion is made in Sec.~\ref{sec:conclusion}.

\section{Boltzmann transport equation}
\label{sec:BTEtheory}

The Boltzmann transport equation for coupled electron and phonon thermal transport is~\cite{IJHMT_2006_EPcoupling_BTE,JHTelectron-phonon2009,JAP2016EP_review}
\begin{align}
\frac{ \partial f_e }{\partial t} + \bm{v}_e \cdot \nabla f_e &= Q_{e-e} + Q_{e-p}+ w_s S_e,  \label{eq:epBTEee}  \\
\frac{ \partial f_p }{\partial t} + \bm{v}_p \cdot \nabla f_p &= Q_{p-p} + Q_{p-e},
\label{eq:epBTEf}
\end{align}
where the subscripts $e$ and $p$ represent electron and phonon, respectively.
Note that this subscript representation will be used throughout the rest of this article.
$f$ is the distribution function, $\bm{v}$ is the group velocity, $S_e$ is the external heat source at the mesoscopic level, $w_s$ is the associated weight satisfying
\begin{align}
\int w_s d\bm{K} =1,
\end{align}
where $ \int d\bm{K}$ represents the integral over the whole first Brillouin zone.
In actual pump-probe experiments, the external heat source originates from the complex and ultrafast photon-electron-phonon coupling process, which is not discussed in detail in this work~\cite{karna_direct_2023,Sciadv_2019_hotelectron}.

The first term on the left-hand side of Eqs.(\ref{eq:epBTEee},\ref{eq:epBTEf}) represents temporal evolution of distribution function and the second term represents electron/phonon advection.
$Q_{e-e}$ and $Q_{p-p}$ represent the electron-electron scattering and phonon-phonon scattering, respectively.
$Q_{e-p}$ and $Q_{p-e}$ are the electron scattering term and phonon scattering term characterizing electron-phonon interaction, respectively.
The actual scattering processes between (quasi) particles are much complicated so that they are reduced in relaxation form in this work,
\begin{align}
Q_{e-e}  &= \frac{ f_e^{eq}  -f_e }{\tau_e } , \\
Q_{p-p}  &= \frac{ f_p^{eq}  -f_p }{\tau_p } ,
\end{align}
where $\tau$ is the associated relaxation time, and $f^{eq}$ is the equilibrium state satisfying the Fermi-Dirac and Bose-Einstein distribution for electron and phonon~\cite{ChenG05Oxford,kittel1996introduction}, respectively,
\begin{align}
f_{e}^{eq}( T_e)  &= \frac{1}{ \exp \left( \frac{\varepsilon - \mu }{k_B T_e} \right) +1 }, \\
f_{p}^{eq}(T_p)  &= \frac{1}{ \exp \left(  \frac{\hbar \omega }{k_B T_p} \right)  -1   },
\end{align}
where $T$ is the temperature, $\varepsilon$ is the electron energy level, $\mu$ is the chemical potential, $k_B$ is the Boltzmann constant, $\hbar$ is the planck constant reduced by $2 \pi$, $\omega$ is the phonon angular frequency.

Equations~\eqref{eq:epBTEee} and~\eqref{eq:epBTEf} can be written as in terms of energy density,
\begin{align}
\frac{ \partial u_e }{\partial t} + \bm{v}_e \cdot \nabla u_e &= \frac{ u_e^{eq}  -u_e }{\tau_e }  - w_g G(T_e -T_p) \notag \\
& + w_s  S,  \label{eq:epBTE1} \\
\frac{ \partial u_p }{\partial t} + \bm{v}_p \cdot \nabla u_p &= \frac{ u_p^{eq}  -u_p }{\tau_p } + w_g G(T_e -T_p),
\label{eq:epBTE2}
\end{align}
where
\begin{align}
u_e-u_e^{eq}(T_{\text{ref} })&= \left(f_e- f_e^{eq}(T_{\text{ref}}) \right)  \left(\varepsilon-E_f\right) D_e ,  \label{eq:equilibrium1}   \\
u_p  -u_p^{eq}(T_{\text{ref} })  &= \left(  f_p- f_p^{eq}(T_{\text{ref} }) \right) \hbar \omega  D_p ,  \label{eq:equilibrium2}  \\
S &= S_e  \left(\varepsilon-E_f\right) D_e ,
\end{align}
where $E_f$ is the Fermi energy, $D$ is the density of states, $T_{\text{ref} }$ is the reference temperature.
We assume $\mu \approx  E_f$ when $T \ll E_f/k_B$.
The electron-phonon coupling process in Eqs.~(\ref{eq:epBTE1},\ref{eq:epBTE2}) is simplified as
\begin{align}
Q_{p-e}=-Q_{e-p}= w_g G(T_e -T_p)
\end{align}
by invoking an electron-phonon coupling parameter $G$, which represents the inelastic scattering that relaxes the electron and phonon to thermal equilibrium.
$w_g$ is the associated weights satisfying $\int w_g d\bm{K} =1$ and the interaction strength between phonons with various modes and electrons with various energy level may be different.
The energy conservation is satisfied during the scattering process so that
\begin{align}
\int \frac{ u_e^{eq}  -u_e }{\tau_e }  d\bm{K} &= 0,  \label{conservation1} \\
\int \frac{ u_p^{eq}  -u_p }{\tau_p }  d\bm{K} &= 0.  \label{conservation2}
\end{align}
The macroscopic variables including the energy $U$ and heat flux $\bm{q}$ are obtained by taking the moment of distribution function,
\begin{align}
U_e  &= \int u_e d\bm{K}, \\
U_p  &= \int u_p d\bm{K}, \\
\bm{q}_e  &= \int \bm{v}_e  u_e d\bm{K}, \\
\bm{q}_p  &= \int \bm{v}_p  u_p d\bm{K}.
\end{align}
The temperature is calculated by following constrictions with Newton iteration method,
\begin{align}
\int u_e^{eq}(T_e) d\bm{K}  &= \int u_e d\bm{K}, \\
\int u_p^{eq}(T_p) d\bm{K}  &= \int u_p d\bm{K}.
\end{align}
Actually in non-equilibrium systems the temperature calculated by above formulas is not the thermodynamic temperature but more like the symbol of local energy density~\cite{ChenG05Oxford}.

To simplify the computation, the frequency-independent assumption is used so that $|\bm{v}|$ and $\tau$ are constants for a given temperature.
The three-dimensional materials are considered and the first Brillouin zone is assumed isotropic just like a spherical surface with fixed radius.
$w_g $ and $w_s$ are both set to be $1 /(4 \pi)$.
The temperature-dependent specific heat is introduced in Eqs.~(\ref{eq:equilibrium1},\ref{eq:equilibrium2})~\cite{JHTelectron-phonon2009},
\begin{align}
u_e-u_e^{eq}(T_{\text{ref} }) &\approx  \frac{1}{4 \pi} \int_{T_{\text{ref} } }^{T_e }   C_e  dT ,  \label{specificheat1}  \\
u_p  -u_p^{eq}(T_{\text{ref} }) &\approx   \frac{1}{4 \pi} \int_{T_{\text{ref} } }^{T_p }   C_p  dT  ,   \label{specificheat2}
\end{align}
where $C_e = \partial \left( f_e^{eq} \left(\varepsilon-E_f\right) D_e \right) / \partial T_e $, $C_p = \partial \left(  f_p^{eq} \hbar \omega  D_p \right) / \partial T_p $.

\section{Discrete unified gas kinetic scheme}
\label{sec:DUGKS}

To solve Eqs.~(\ref{eq:epBTE1},\ref{eq:epBTE2}), the discrete unified gas kinetic scheme (DUGKS)~\cite{guo_progress_DUGKS,PhysRevE.107.025301} is used, which has made great success in multiscale particle transport~\cite{guo_progress_DUGKS}.
The whole time space, spatial space and first Brillouin zone are discretized into many small pieces, and Eqs.~(\ref{eq:epBTE1},\ref{eq:epBTE2}) in integral form over a control volume $i$ from time $t_m$ to $t_{m+1}=t_{m}+ \Delta t$ can be written as follows,
\begin{align}
&u_{e,i,n}^{m+1}-u_{e,i,n}^{m} + \frac{\Delta t}{V_i} \sum_{j \in N(i)} \left(  \bm{v}_e \cdot \mathbf{n}_{ij} u_{e,ij,n}^{m+1/2} A_{ij} \right)  \notag \\
 &=\frac{\Delta t}{2}  \left( H_{i,n}^{m+1}  + H_{i,n}^{m} \right),  \label{eq:dpBTE1} \\
&u_{p,i,n}^{m+1}-u_{p,i,n}^{m} + \frac{\Delta t}{V_i} \sum_{j \in N(i)} \left(  \bm{v}_p \cdot \mathbf{n}_{ij} u_{p,ij,n}^{m+1/2} A_{ij} \right)  \notag \\
 &=\frac{\Delta t}{2}  \left( F_{i,n}^{m+1} + F_{i,n}^{m}  \right),  \label{eq:dpBTE2}
\end{align}
where the trapezoidal quadrature is used for the time integration of the scattering, electron-phonon coupling and heat source terms, while the mid-point rule is used for the flux term. $H= (u_e^{eq}  -u_e )/ \tau_e - w_g G(T_e -T_p) + w_s S$, $F=( u_p^{eq}  -u_p ) /  \tau_p  + w_g G(T_e -T_p)$, $n$ represents the index of discretized first Brillouin zone, $V_i$ is the volume of the cell $i$, $N(i)$ denotes the sets of neighbor cells of cell $i$, $ij$ denotes the interface between cell $i$ and cell $j$, $A_{ij}$ is the area of the interface $ij$, $\mathbf{n}_{ij}$ is the normal unit vector of the interface $ij$ directing from cell $i$ to cell $j$, $\Delta t$ is the time step and $m$ is an index of time step.
Reformulate above two equations~(\ref{eq:dpBTE1},\ref{eq:dpBTE2}) as
\begin{align}
\tilde{I}_{e,i,n}^{m+1}  &= \tilde{I}_{e,i,n}^{+,m} -   \frac{\Delta t}{V_i} \sum_{j \in N(i)} \left(  \bm{v}_e \cdot \mathbf{n}_{ij} u_{e,ij,n}^{m+1/2} A_{ij} \right),  \label{dbte3}  \\
\tilde{I}_{p,i,n}^{m+1}  &= \tilde{I}_{p,i,n}^{+,m} -  \frac{\Delta t}{V_i} \sum_{j \in N(i)} \left(  \bm{v}_p \cdot \mathbf{n}_{ij} u_{p,ij,n}^{m+1/2} A_{ij} \right)  \label{dbte4},
\end{align}
where
\begin{align}
\tilde{I}_{e}   &= u_{e} - \Delta t   H /2 ,  \label{eq:Iecenters}    \\
\tilde{I}_{e}^{+} &= u_{e}  + \Delta t H /2 ,   \\
\tilde{I}_{p}  &= u_{p} - \Delta t   F /2 , \label{eq:Ipcenters}   \\
\tilde{I}_{p}^{+} &= u_{p}  + \Delta t F /2  .
\end{align}

In order to obtain the distribution function at the cell interface at the mid-point time step ($u_{e,ij,n}^{m+1/2}$, $u_{p,ij,n}^{m+1/2}$), integrating Eqs.~(\ref{eq:epBTE1},\ref{eq:epBTE2}) from time $t_m$ to $t_{m+1/2}=t_{m}+ \Delta t/2$ along the characteristic line with the end point $\bm{x}_{ij}$ locating at the center of the cell interface $ij$ between cell $i$ and cell $j$,
\begin{align}
&u_{e}^{m+1/2} (\bm{x}_{ij} ) -u_{e}^{m} (\bm{x}_{ij} -\bm{v}_e \Delta t/2 ) \notag \\
 &= \Delta t/4 \left( H^{m+1/2}(\bm{x}_{ij} ) +H^{m }(\bm{x}_{ij} -\bm{v}_e \Delta t/2 )  \right) , \label{eq:BTEfaces1}   \\
&u_{p}^{m+1/2} (\bm{x}_{ij} ) -u_{p}^{m} (\bm{x}_{ij} -\bm{v}_p \Delta t/2 ) \notag \\
 &= \Delta t/4 \left( F^{m+1/2}(\bm{x}_{ij} ) +F^{m }(\bm{x}_{ij} -\bm{v}_p \Delta t/2 )  \right).  \label{eq:BTEfaces2}
\end{align}
Reformulate above two equations as
\begin{align}
\bar{I}_{e}^{m+1/2}(\bm{x}_{ij}) &= \bar{I}_{e}^{+,m}(\bm{x}_{e,ij}') ,  \\
\bar{I}_{p}^{m+1/2}(\bm{x}_{ij}) &= \bar{I}_{p}^{+,m}(\bm{x}_{p,ij}') ,
\end{align}
where
\begin{align}
\bar{I}_{e}  &= u_{e} - \Delta t   H /4 ,  \label{eq:Iefaces}    \\
\bar{I}_{e}^{+} &= u_{e}  + \Delta t H /4 ,   \\
\bar{I}_{p} &= u_{p} - \Delta t   F /4 , \label{eq:Ipfaces}   \\
\bar{I}_{p}^{+} &= u_{p}  + \Delta t F /4  ,   \\
\bm{x}_{e,ij}' &= \bm{x}_{ij} -\bm{v}_e \Delta t/2 , \\
\bm{x}_{p,ij}' &= \bm{x}_{ij} -\bm{v}_p \Delta t/2 .
\end{align}
$\bar{I}_{e}^{+,m}(\bm{x}_{e,ij}')$ and $\bar{I}_{p}^{+,m}(\bm{x}_{p,ij}')$ are reconstructed by numerical interpolation,
\begin{align}
\bar{I}_{e}^{+,m}(\bm{x}_{e,ij}') = \bar{I}_{e}^{+,m} (\bm{x}_{c}) + (\bm{x}_{e,ij}'-\bm{x}_{c}) \bm{\sigma}_{e,c},  \\
\bar{I}_{p}^{+,m}(\bm{x}_{p,ij}') = \bar{I}_{p}^{+,m} (\bm{x}_{c}) + (\bm{x}_{p,ij}'-\bm{x}_{c}) \bm{\sigma}_{p,c},
\label{eq:slope}
\end{align}
where $\bm{\sigma}_{c}$ is the spatial gradient of the distribution function $\bar{I}^{+}$ at the cell $c$.
If $ \bm{v}  \cdot \bm{n}_{ij} >0$, $c=i$; else $c=j$.
The first-order upwind scheme, van Leer limiter or least square method can be adopted to calculate the spatial gradient to ensure the numerical stability.

Once $\bar{I}_{e}^{m+1/2}$ and $\bar{I}_{p}^{m+1/2}$ are obtained, taking an integral of Eqs.~(\ref{eq:Iefaces},\ref{eq:Ipfaces}) over the whole first Brillouin zone leads to
\begin{align}
\frac{ \Delta t}{4}  S + \int  \bar{I}_{e}  d\bm{K} &=  U_e  + \frac{ \Delta t}{4} G(T_e -T_p)  ,  \\
\int    \bar{I}_{p} d\bm{K} &=  U_p  - \frac{ \Delta t}{4} G(T_e -T_p) .
\end{align}
Combining Eqs.~(\ref{specificheat1},\ref{specificheat2},\ref{conservation1},\ref{conservation2}) and above two equations, the electron and phonon temperatures at the cell interface can be calculated by Newton iteration method~\cite{zhang_discrete_2019}.
Then the original distribution function $u_e^{m+1/2}$ and $u_p^{m+1/2}$ at the cell interface can be obtained based on Eqs.~(\ref{eq:Iefaces},\ref{eq:Ipfaces}).
Similar treatment is conducted for the evolutions of the distribution function and macroscopic variables at the cell center.
$\tilde{I}_{e}$ and $\tilde{I}_{p}$ can be updated based on Eqs.~(\ref{dbte3},\ref{dbte4}), then taking an integral of Eqs.~(\ref{eq:Iecenters},\ref{eq:Ipcenters}) over the whole first Brillouin zone leads to
\begin{align}
\frac{ \Delta t}{2}  S + \int  \tilde{I}_{e}    d\bm{K} &=  U_e  + \frac{ \Delta t}{2} G(T_e -T_p)  ,  \\
\int    \tilde{I}_{p} d\bm{K} &=  U_p  - \frac{ \Delta t}{2} G(T_e -T_p) .
\end{align}
Combining Eqs.~(\ref{specificheat1},\ref{specificheat2},\ref{conservation1},\ref{conservation2}) and above two equations, the electron and phonon temperatures at the cell center can be calculated by Newton iteration method.
Then the original distribution function $u_e^{m+1}$ and $u_p^{m+1}$ at the cell interface can be updated.

Above procedures are the main evolution processes of the phonon and electron distribution function in the DUGKS.
Actually the numerical evolutions process of DUGKS is not limited to frequency-independent and isotropic assumptions. Related extension will be conducted in the future.
The key difference between the DUGKS and typical Monte Carlo method or discrete ordinate method is the reconstruction of the distribution function at the cell interface.
Instead of direct numerical interpolation, the BTE is solved again at the cell interface (Eqs.~\ref{eq:BTEfaces1},\ref{eq:BTEfaces2}) so that the physical evolution process is included adaptively and the electron/phonon advection, scattering and electron-phonon interactions are coupled together within one time step~\cite{PhysRevE.107.025301,guo_progress_DUGKS}.

\section{RESULTS AND DISCUSSIONS}
\label{sec:results}

In this section, numerical simulations are conducted and the present results are compared to the experimental data or results predicted by macroscopic two-temperature model (TTM, see appendix~\ref{sec:TTM}).
The time step of DUGKS is $\Delta t= \text{CFL} \times  \Delta x / v_{max}   $, where $0< \text{CFL} <1 $ is the Courant–Friedrichs–Lewy number, $\Delta x$ is the minimum cell size and $v_{max}$ is the maximum group velocity of electron and phonon.
In the following simulations, $\text{CFL}=0.4 $ or $0.8$.
The first-order upwind scheme is used in the ballistic regime, and van Leer limiter is used in the diffusive regime.

The remainder of this section is organized as follows.
The first three subsections are mainly to verify the effectiveness of the present scheme.
In subsection~\ref{sec:laser-heating}, the DUGKS results are compared with the experimental data in existing references and the TTM.
All input parameters and experimental data are obtained from the existing references.
In subsection~\ref{sec:TDTR}, a time-domain thermoreflectance (TDTR) experimental platform is used to measure the reflected signals, and the experimental data is fitted by the DUGKS to get a reasonable electron-phonon coupling constant.
In subsection~\ref{sec:cross_plane}, the steady quasi-1D cross-plane heat conduction is simulated and results show that the present scheme could predict the thermal transport accurately even when the cell size is much larger than the mean free path.
In subsection~\ref{sec:TTG}, the dynamics of electron and phonon in transient thermal grating geometry is studied by DUGKS, and the thermal behaviors and differences in the ballistic and diffusive regimes are discussed.
In the final subsection~\ref{sec:bilayer}, the heat conduction in bilayer metals is studied accounted for interfacial thermal resistance.

\subsection{Ultrafast laser heating for electron-phonon coupling constant}
\label{sec:laser-heating}

\begin{figure}[htb]
\centering
\includegraphics[scale=0.6,clip=true]{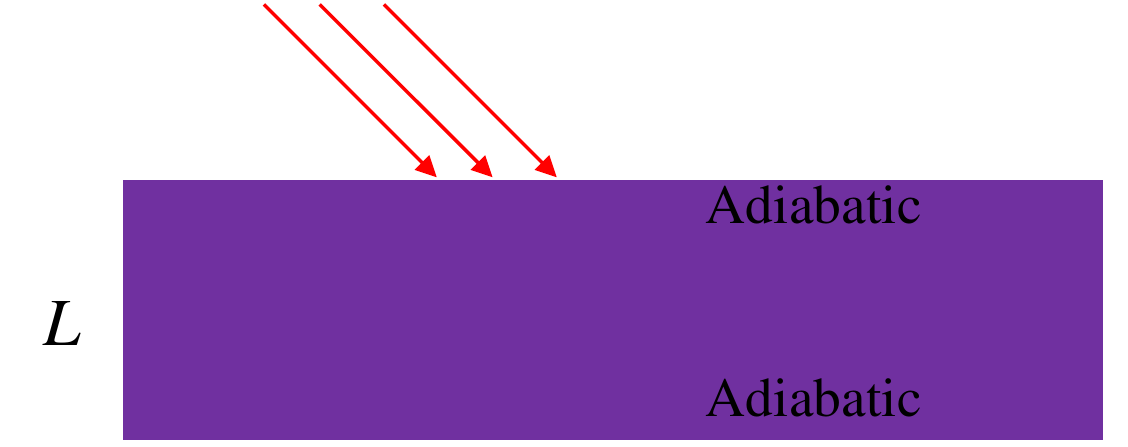}
\caption{Ultrafast laser heating of single-layer metals.}
\label{Au_film_schematic}
\end{figure}
\begin{figure*}[htb]
\centering
\subfloat[$L=100$ nm, front surface]{ \includegraphics[scale=0.3,clip=true]{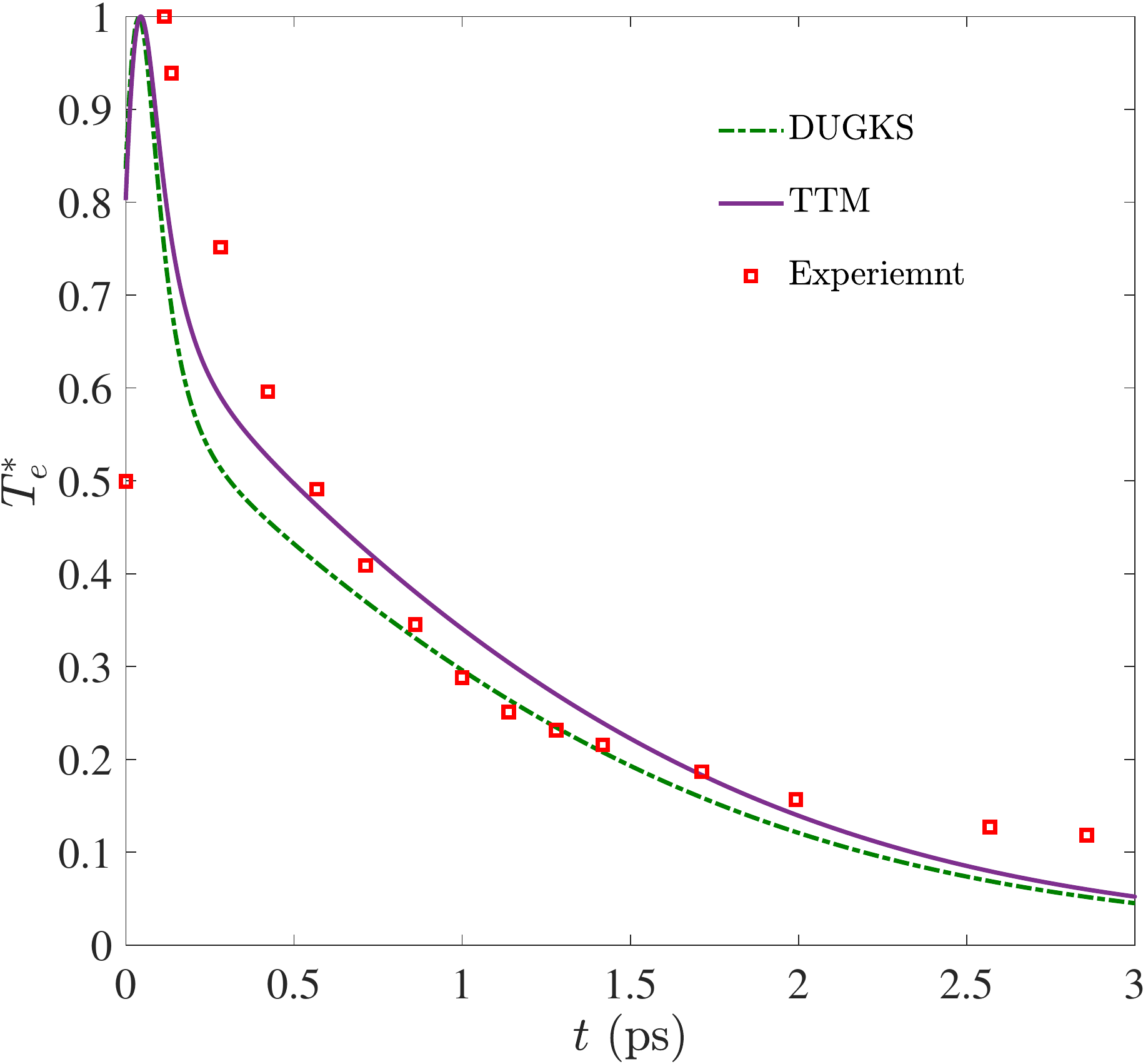} }  ~~
\subfloat[$L=100$ nm, rear surface]{ \includegraphics[scale=0.3,clip=true]{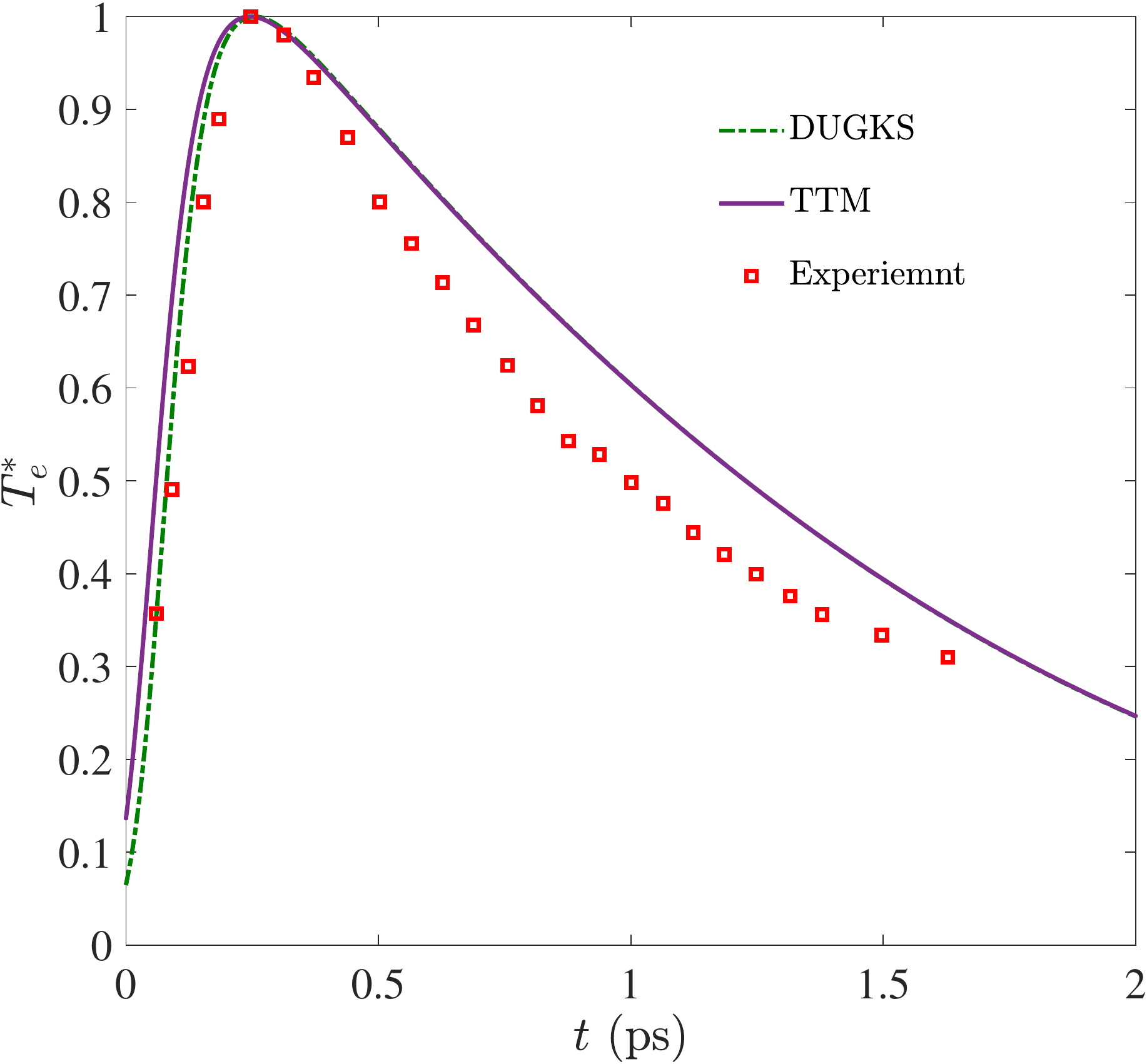} }  \\
\subfloat[$L=200$ nm, front surface]{ \includegraphics[scale=0.3,clip=true]{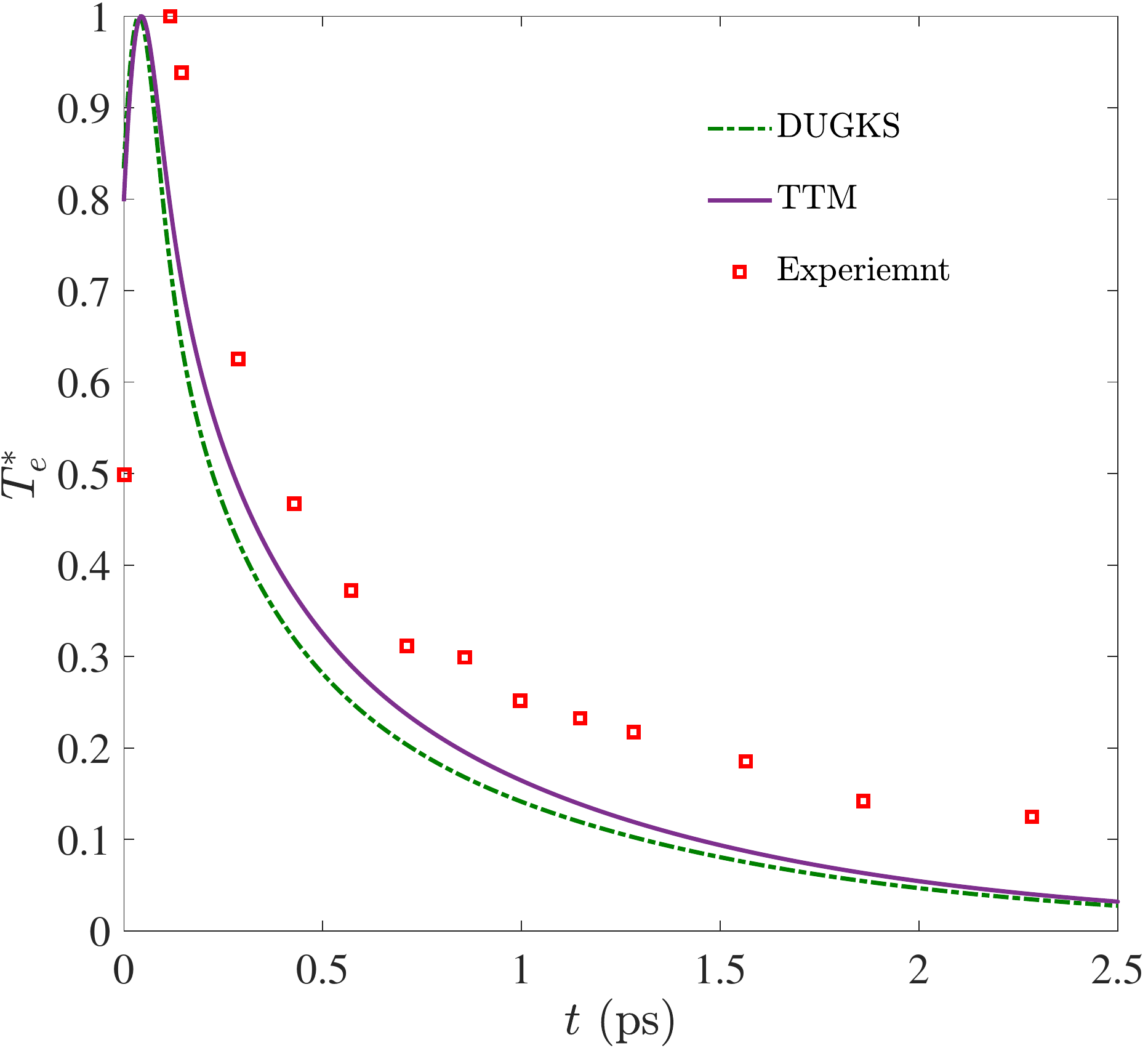} }  ~~
\subfloat[$L=200$ nm, rear surface]{ \includegraphics[scale=0.3,clip=true]{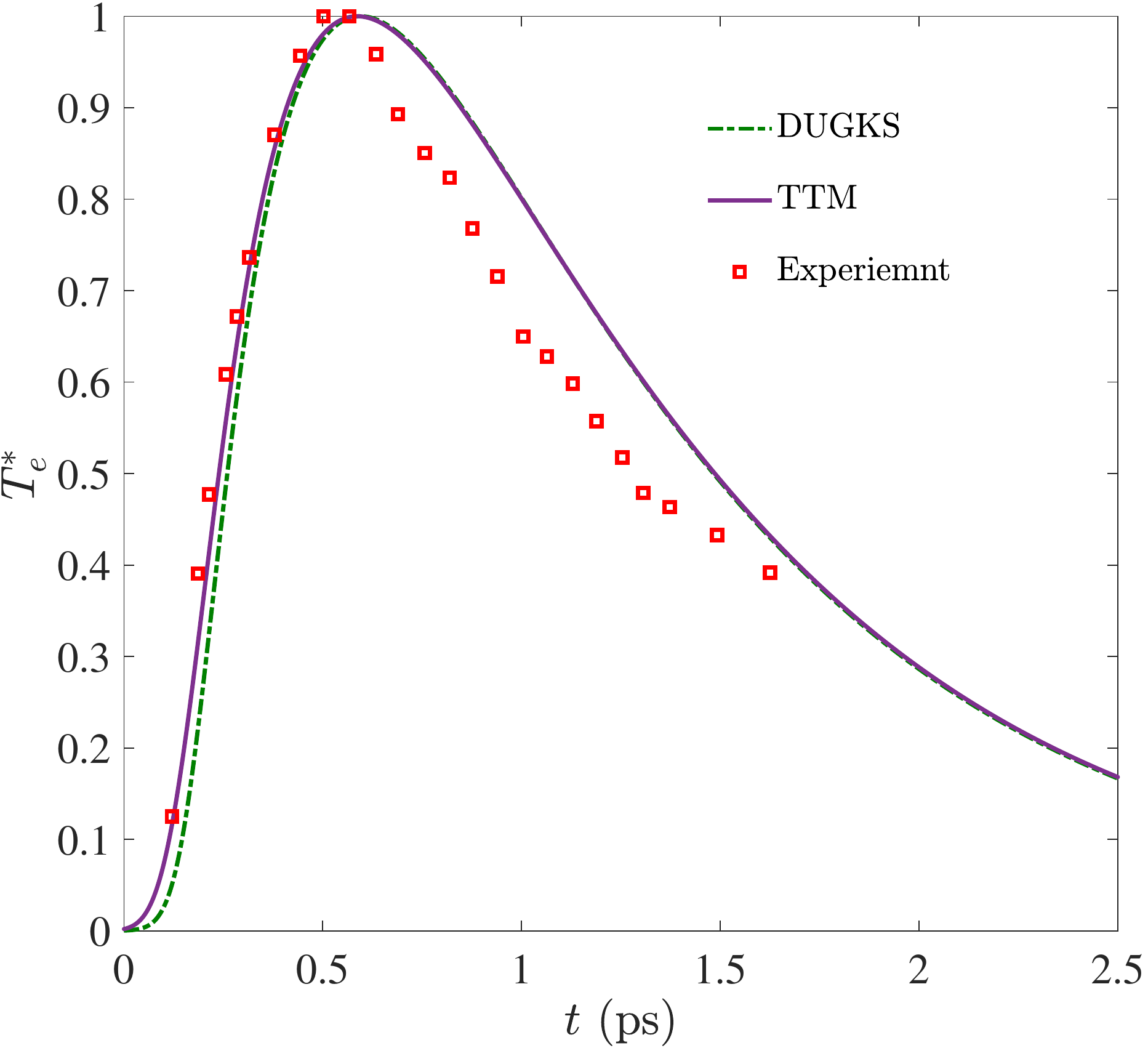} }  \\
\caption{Comparison of the normalized electron temperature predicted by the present DUGKS results, two-temperature model (TTM) and experiments~\cite{PhysRevLett.59.1962,JHTelectron-phononTienCL} for single-layer Au film, where $T_e^*= (T_e - T_{\text{ref}} )/(T_{max} - T_{\text{ref}} )$ and $T_{max}$ is the maximum electron temperature.}
\label{Au_film_results}
\end{figure*}
\begin{figure*}[htb]
\centering
\subfloat[$P_{input} =17.6$ J$\cdot$m$^{-2}$]{ \includegraphics[scale=0.3,clip=true]{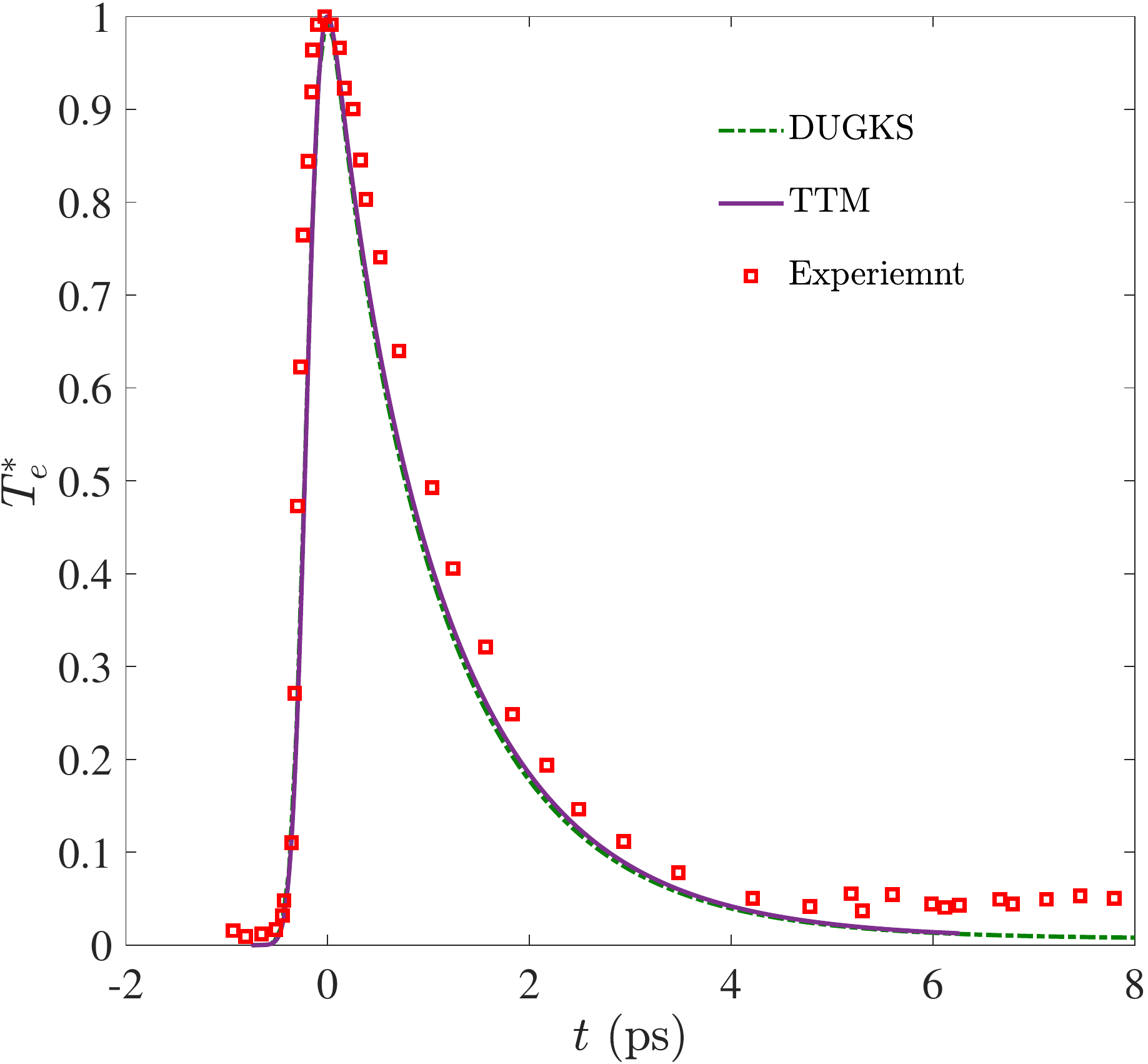} }  ~~
\subfloat[$P_{input} =70.6$ J$\cdot$m$^{-2}$]{ \includegraphics[scale=0.3,clip=true]{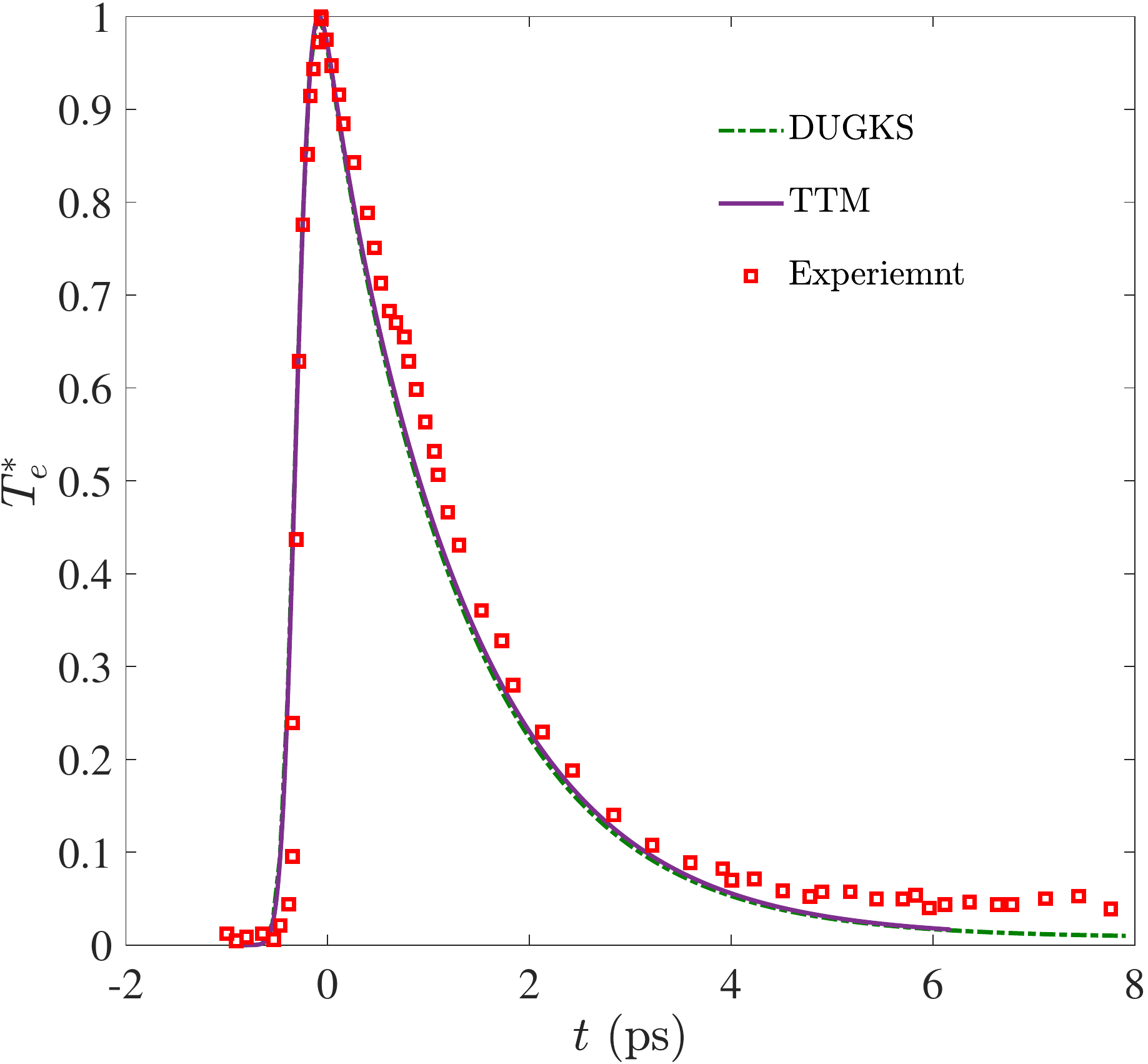} }  \\
\caption{Comparison of the normalized electron temperature predicted by the present DUGKS results, two-temperature model (TTM) and experiments~\cite{Liang_JHT2014} for single-layer Au film with fixed thickness $L=~1\mu$m and various laser energy input $P_{input}$, where $T_e^*= (T_e - T_{\text{ref}} )/(T_{max} - T_{\text{ref}} )$ and $T_{max}$ is the maximum electron temperature.}
\label{Au_film_pulseinput}
\end{figure*}

The quasi-1D thermal transport of the single-layer Au films with different thicknesses $L$ was studied, as shown in~\cref{Au_film_schematic}.
The initial temperature in the whole domain is $T_{\text{ref}}=300$ K and the ultrafast laser heating is implemented on the front surface $x=0$,
\begin{align}
S=  \frac{P_{input} (1- R_{r} )  }{ t_{s} d_{pump}  }  \exp{ \left( -\frac{x}{d_{pump} } -  \frac{ (4 \ln{2} ) t^2}{t_{s}^2 }    \right)  },
\label{eq:sourcepump}
\end{align}
where $R_{r} $ is the optical reflectivity, $d_{pump}$ is the optical penetration depth, $P_{input}$ is the total energy carried by a laser pulse divided by the laser spot cross section, $t_s=t_{pump}+t_h $, $t_h$ is the delay in the electron thermalization time after pulse absorption~\cite{Nonlinear_Thermore2011JHT,PhysRevB.46.13592,PhysRevB.48.12365,HOHLFELD2000237} after pulse absorption, $t_{pump}$ is the full-width-at-half-maximum (FWHM) duration of the laser pulse.

Thermal physical parameters of electrons in Au metals are listed below: specific heat is $C_e = \gamma T_e$, $\gamma=71$ J$\cdot$m$^{-3}$$\cdot$K$^{-2}$, group velocity is $| \bm{v}_e |= 1.36 \times 10^6$ m$\cdot$s$^{-1}$,  $E_f = 5.51$ eV, thermal conductivity is~\cite{JHTelectron-phonon2009,anisimov1997theory}
\begin{align}
\kappa_e = \chi \frac{(\vartheta_e^2 +0.16)^{5/4} (\vartheta_e^2 +0.44 ) \vartheta_e }{ (\vartheta_e^2 +0.092 )^{51/2}  (\vartheta_e^2 + \eta \vartheta_p  )  },
\end{align}
where $\chi = 353$ W$\cdot$m$^{-1}$$\cdot$K$^{-1}$, $\eta=0.16$, $\vartheta_e =k_B T_e /E_f$, $\vartheta_p =k_B T_p /E_f$.
The relaxation time is calculated by the kinetic relation $\tau_e = 3 \kappa_e/(C_e |\bm{v}_e|^2 )$.
Thermal physical parameters of phonons in Au metals are listed below: specific heat is $C_p =2.5 \times 10^6$ J$\cdot$m$^{-3}$$\cdot$K$^{-1}$,
group velocity is $| \bm{v}_p |= 2142.86 $ m$\cdot$s$^{-1}$, thermal conductivity is $\kappa_p = 2.6$ W$\cdot$m$^{-1}$$\cdot$K$^{-1}$, relaxation time is $\tau_p = 3 \kappa_p/(C_p |\bm{v}_p|^2 )$.
The phonon and electron mean free path are about $1.5$ nm and $33$ nm, respectively.

The transient heat conduction in single-layer Au film with different thickness $L$ is simulated by DUGKS, and the dynamics of electron temperature at the front $x=0$ and rear $x=L$ surfaces are plotted.
In these simulations~\cite{MOZAFARIFARD2023123759,JHTelectron-phononTienCL}, $t_{pump}= 96$ fs, $t_h=0$ fs, $d_{pump}=15.3$ nm, $R_{reflection}=0.93$, $P_{input} =10$ J$\cdot$m$^{-2}$, $G=2.6\times 10^{16}$ W$\cdot$m$^{-3}$$\cdot$K$^{-1}$.
The computational domain is discretized with $20-80$ uniform cells, and $40$ discrete points in the $|\bm{v}| \cos \theta $ direction is used to ensure the accuracy of the numerical quadrature, where $\theta \in [0, \pi]$.
The specular reflection boundary conditions (Eq.~\eqref{eq:BC2}) are adopted for both the front and rear surfaces.
The evolutions of normalized electron temperature $T_e^*= (T_e - T_{\text{ref}} )/(T_{max} - T_{\text{ref}} )$ are shown in~\cref{Au_film_results}, where $T_{max}$ is the maximum electron temperature and we assume that the change in the reflected signal is linear to the change in electron temperature~\cite{JHTelectron-phononTienCL,MOZAFARIFARD2023123759}.
It can be found that the present DUGKS results remain broadly consistent with the TTM and previous experiments~\cite{PhysRevLett.59.1962,JHTelectron-phononTienCL}.

The deviations between different models and experiments result from following reasons. 1) Uncertainty of experimental parameters~\cite{Liang_JHT2014,JHTelectron-phononTienCL,sivan_ultrafast_2020,PhysRevB.48.12365}, for example, it is very difficult to precisely determine the practical FWHM or optical reflectivity of laser heating, which significantly influence the maximum electron temperature rise. 2) The TTM assumes diffusive electron/phonon transport with infinite heat propagation speed, which may be invalid when the system characteristic length/time is comparable to or smaller than the mean free path/relaxation time.

We also made a comparison with the experimental data measured by Guo $et~ al.$~\cite{Liang_JHT2014} with fixed thickness $L=1~\mu$m and various laser heating pump $P_{input}$.
The electron ballistic depth length $d_{ballistic}$ is introduced and the external heat source is
\begin{align}
S= &P_{input}   \frac{0.94( 1- R_r  )  }{ t_{s} d_{length}  \left( 1- \exp (-L/d_{length} )  \right)   }  \notag \\
 &\times \exp{ \left( -\frac{x}{d_{length} } - (4 \ln{2} )  \frac{t^2}{t_{s}^2 }    \right)  },
\end{align}
where $t_{s}= 280$ fs, $d_{length}=d_{pump}+d_{ballistic}$, $d_{pump}=12.44$ nm, $d_{ballistic}=200$ nm, $R =0.970$.
Thermal physical parameters of electrons and phonon are the same as above except $\kappa_p = 0.311 $ W$\cdot$m$^{-1}$$\cdot$K$^{-1}$ and $G=1.5 \times 10^{16}$ W$\cdot$m$^{-3}$$\cdot$K$^{-1}$.

The evolutions of normalized electron temperature $T_e^*$ are shown in~\cref{Au_film_pulseinput} with different laser heating input $P_{input}$, where the variations of measured reflection signals are approximately linear to the electron temperature increment~\cite{Liang_JHT2014}.
When $L=1~\mu$m, both electron and phonon suffer a diffusive transport process, and the profiles show that the DUGKS results are in excellent agreement with the TTM and experiments.
When $P_{input} =17.6$ J$\cdot$m$^{-2}$, the maximum electron temperature predicted by the DUGKS and TTM are $370$ K and $369.0$ K, respectively.
When $P_{input} =70.6$ J$\cdot$m$^{-2}$, the maximum electron temperature predicted by the DUGKS and TTM are $533$ K and $527.4$ K, respectively.
Those results are in consistent with the data shown in Ref.~\cite{Liang_JHT2014}.

\subsection{Time-domain thermoreflectance experiments}
\label{sec:TDTR}

\begin{figure*}[htb]
\centering
\subfloat[$t_h=0$]{\includegraphics[scale=0.3,clip=true]{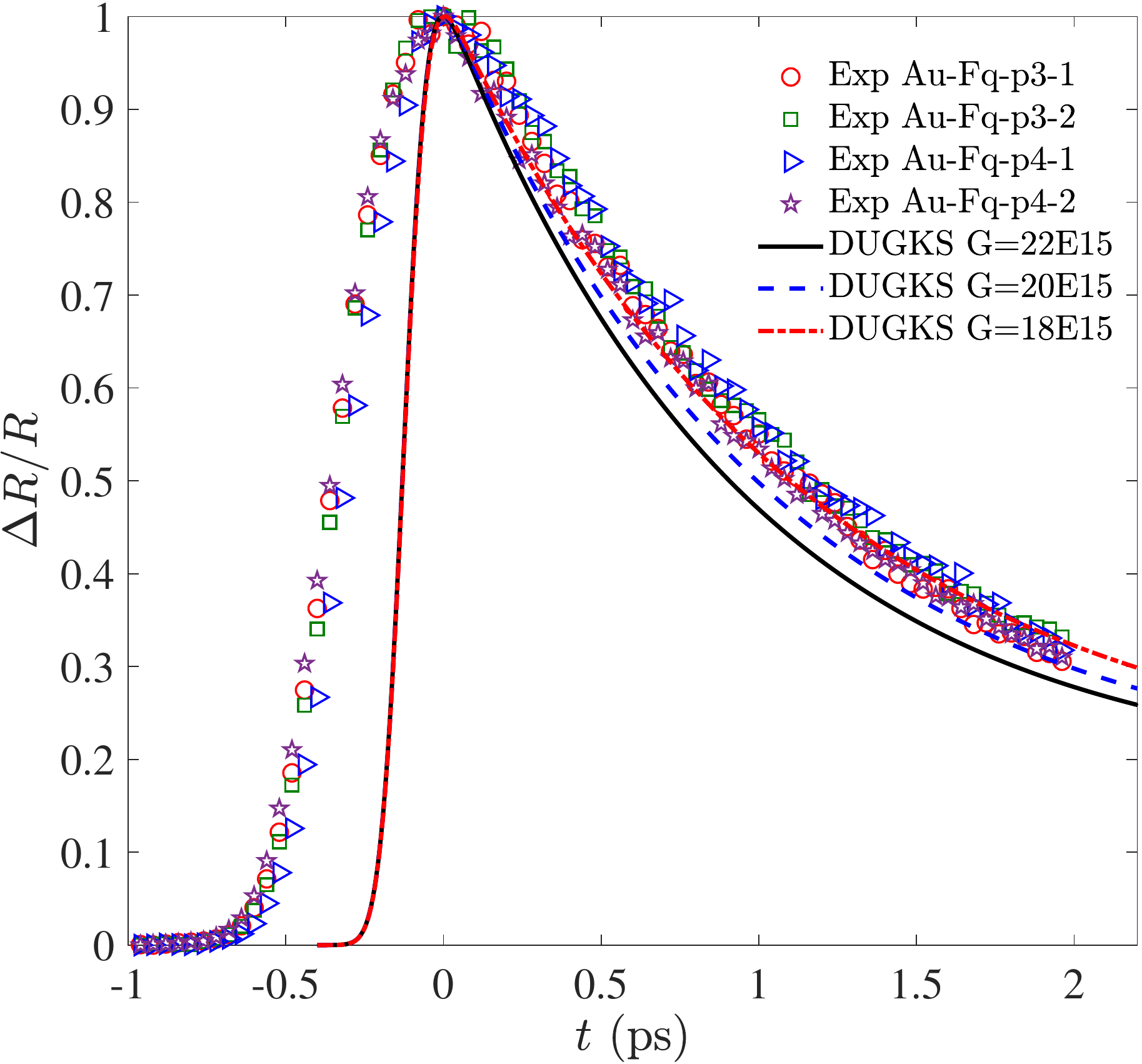} }  ~~
\subfloat[$t_h=260$ fs]{\includegraphics[scale=0.3,clip=true]{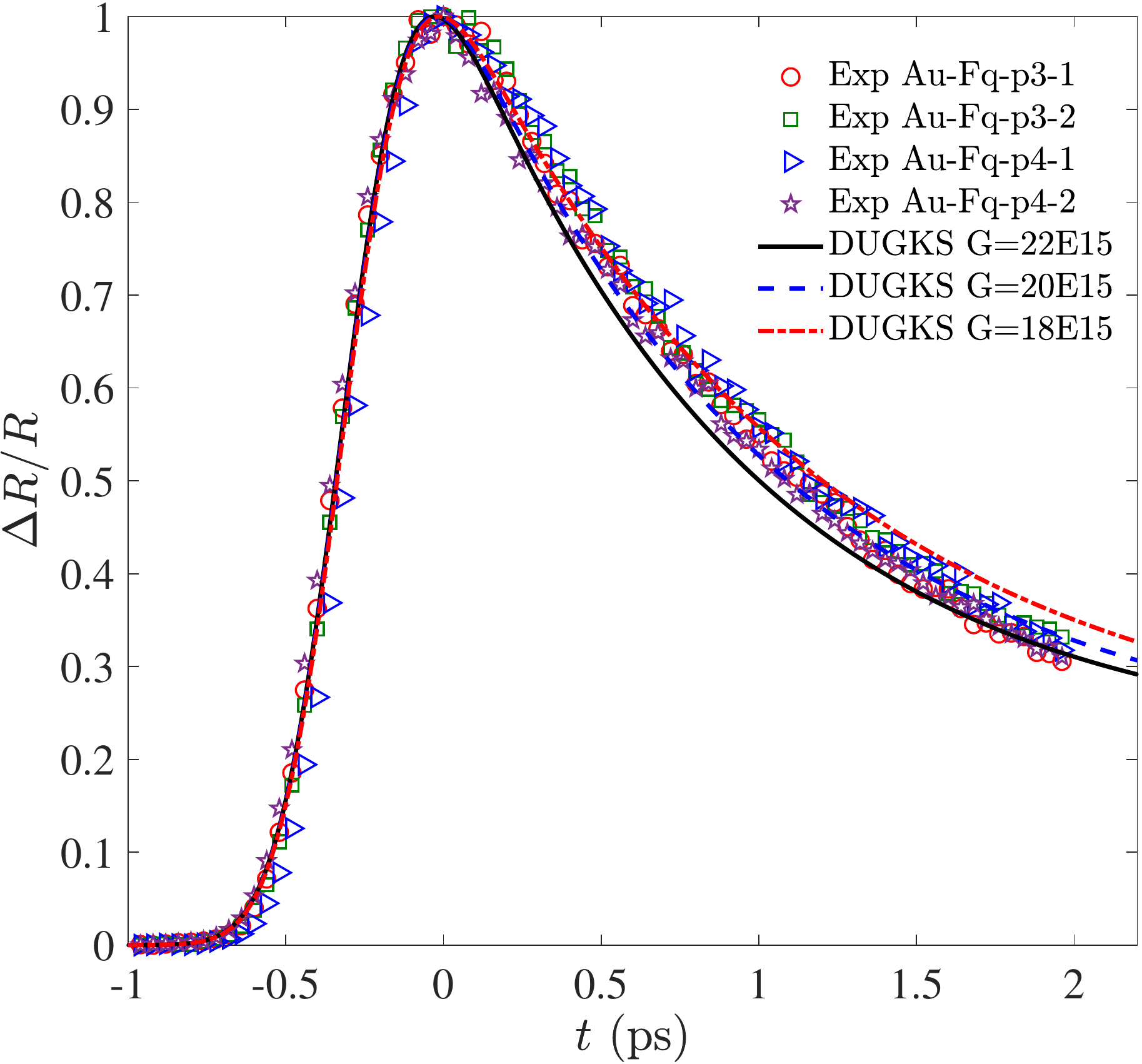} }  \\
\caption{Comparison of the normalized reflected signal $\Delta R/R$ between the TDTR experiments and DUGKS with various electron-phonon coupling constant $G$ and delay in the electron thermalization time after pulse absorption $t_h$~\cite{PhysRevB.46.13592,PhysRevB.48.12365}.}
\label{Au_film_TDTR}
\end{figure*}
The transient reflected signal for Au thin film with thickness $18.2$ nm at environment temperature $T_{\text{ref}} =288.15$ K was measured using a time-domain thermoreflectance (TDTR) experimental platform, which is a widely used thermal property measurement method based on pump-probe technology.
In this method, a modulated pulsed pump beam heats the surface periodically and a delayed pulsed probe beam detects the variation in thermoreflectance signal, which is a function of the temperature of electron and phonon.
The signal picked up by a photodiode and a lock-in amplifier is fitted with an analytical heat conduction solution, such as two-temperature model and Boltzmann’s transport equation. A Ti:Sapphire femtosecond laser (Chameleon II from Coherent Inc.) is utilized to generate pulsed beam with a pulse width of $140$ fs.
The wavelength of pump beam and probe beam are $400$ nm and $800$ nm, respectively.
These two beams are focused on the surface of sample by a $10$x objective lens with diameters of $39.2~\mu$m and $10.9~\mu$m, respectively.
The delay time between probe beam and pump beam is controlled by a motorized linear stage (PRO165SL(E)-600 from Aerotech Inc.) and four-fold light path, which provides a minimum delay time step of $0.01$ ps.

The thermal physical parameters of electrons and phonons are the same as those used in~\cref{Au_film_results} as well as the heat source~\eqref{eq:sourcepump} except $t_{pump}= 140$ fs and $P_{input} (1-R_r ) =0.0112 $ J$\cdot$m$^{-2}$.
The heating spot radius is much larger than the film thickness so that the heat conduction can be approximated as a quasi-1D transient heat conduction problem.
The normalized reflected signal $\Delta R/R$ predicted by DUGKS with fitted electron-phonon coupling constant $G$ and the delay in the electron thermalization time after pulse absorption $t_h$~\cite{Nonlinear_Thermore2011JHT,PhysRevB.46.13592,IBRAHIM20042261,IJHMT_2006_EPcoupling_BTE} are shown in~\cref{Au_film_TDTR}, where the relationship between the reflected signals and electron/phonon temperature is introduced in Appendix~\ref{sec:reflectedsignal}.
It can be found that when $t_h=260$ fs and $G \approx 2.0 \times 10^{16}$ W$\cdot$m$^{-3}$$\cdot$K$^{-1}$, the numerical results are in excellent agreement with the four group TDTR measured data.

There are many uncertainties compared to existing references: 1) different experimental samples and external heat source, 2) different empirical formulas and coefficients of the thermal physical parameters, for example the temperature-dependent electron specific heat.
Hence it can be found that the measured electron-phonon coupling constant of Au in various references are different.
The electron-phonon coupling coefficient for Au thin film measured by our performed TDTR is basically consistent with previous experimental data~\cite{Liang_JHT2014,PhysRevLett.59.1962,JHTelectron-phononTienCL,sivan_ultrafast_2020,PhysRevB.48.12365} ranging from $1.5 \times 10^{16}$ to $4.0 \times 10^{16}$ W$\cdot$m$^{-3}$$\cdot$K$^{-1}$.

\subsection{Steady cross-plane heat conduction}
\label{sec:cross_plane}

\begin{figure}[htb]
\centering
\includegraphics[scale=0.30,clip=true]{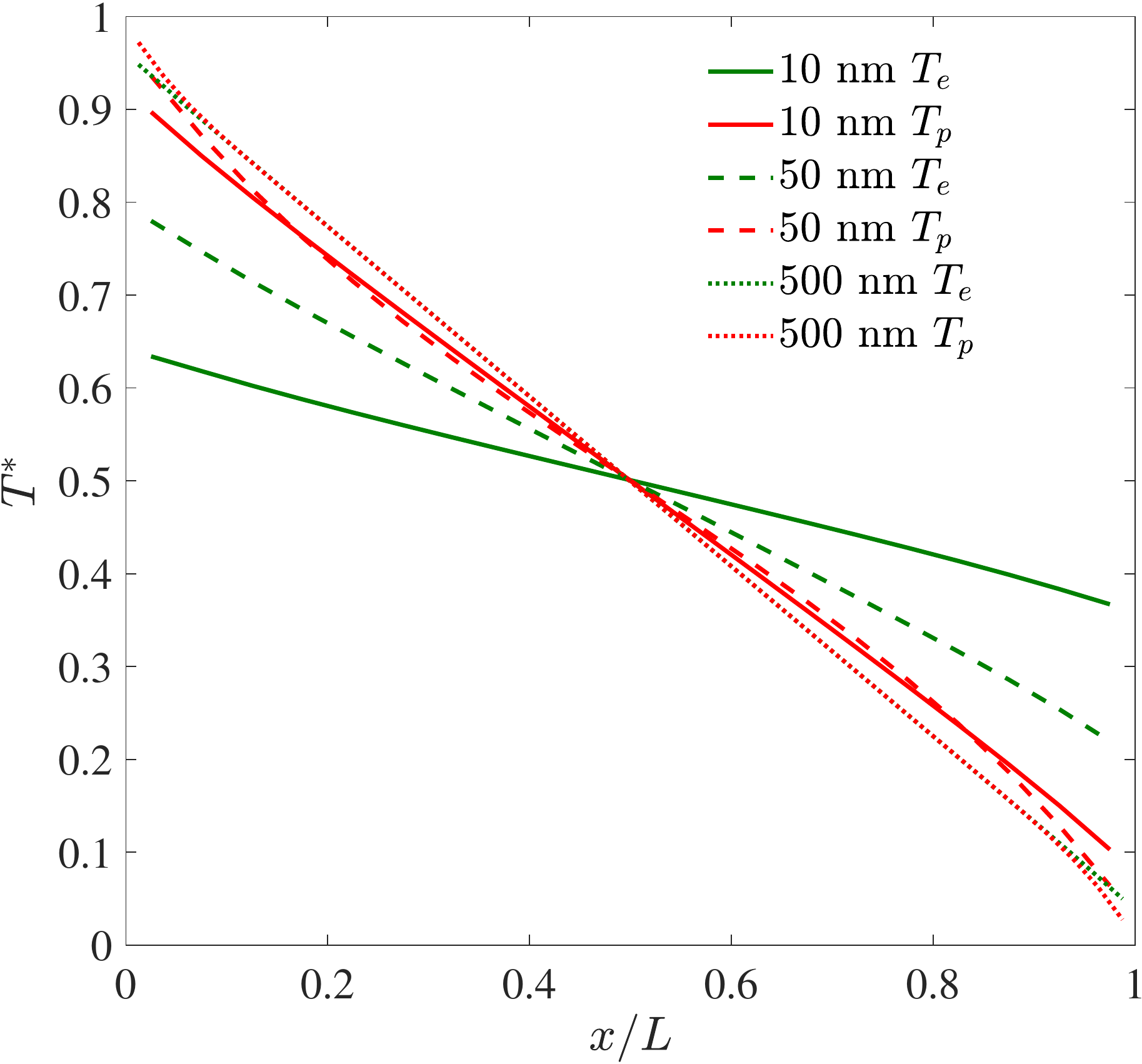}
\caption{Spatial distributions of the electron and phonon temperatures with different thickness $L$, where $T^*=(T-T_R)/(T_L -T_R)$, the red line is the phonon temperature $T_p$ and the dark green is the electron temperature $T_e$. }
\label{Au_cross_plane}
\end{figure}
\begin{figure}[htb]
\centering
\subfloat[$L=100$ nm]{ \includegraphics[scale=0.30,clip=true]{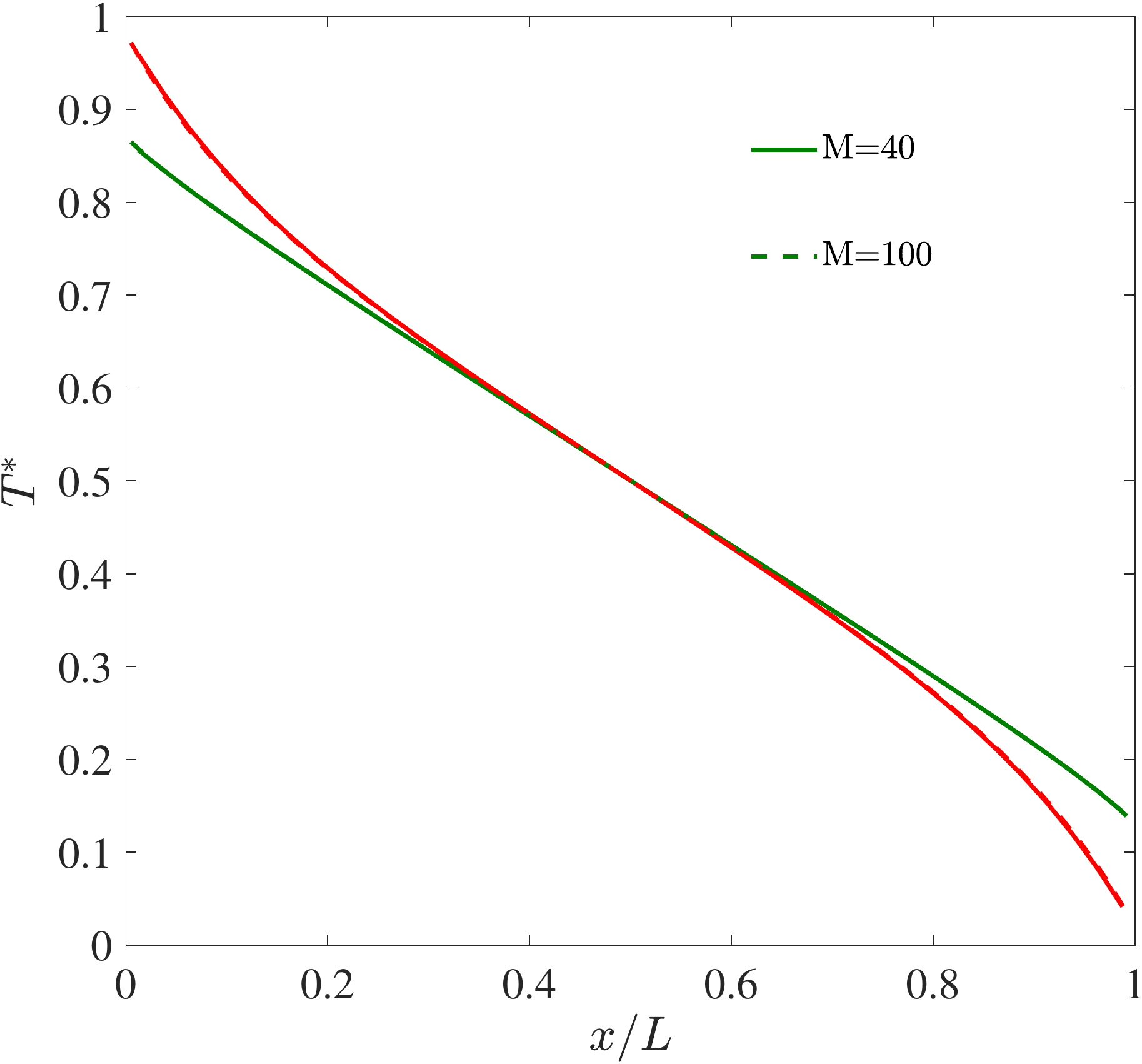} } \\
\subfloat[$L=10~\mu$m]{ \includegraphics[scale=0.30,clip=true]{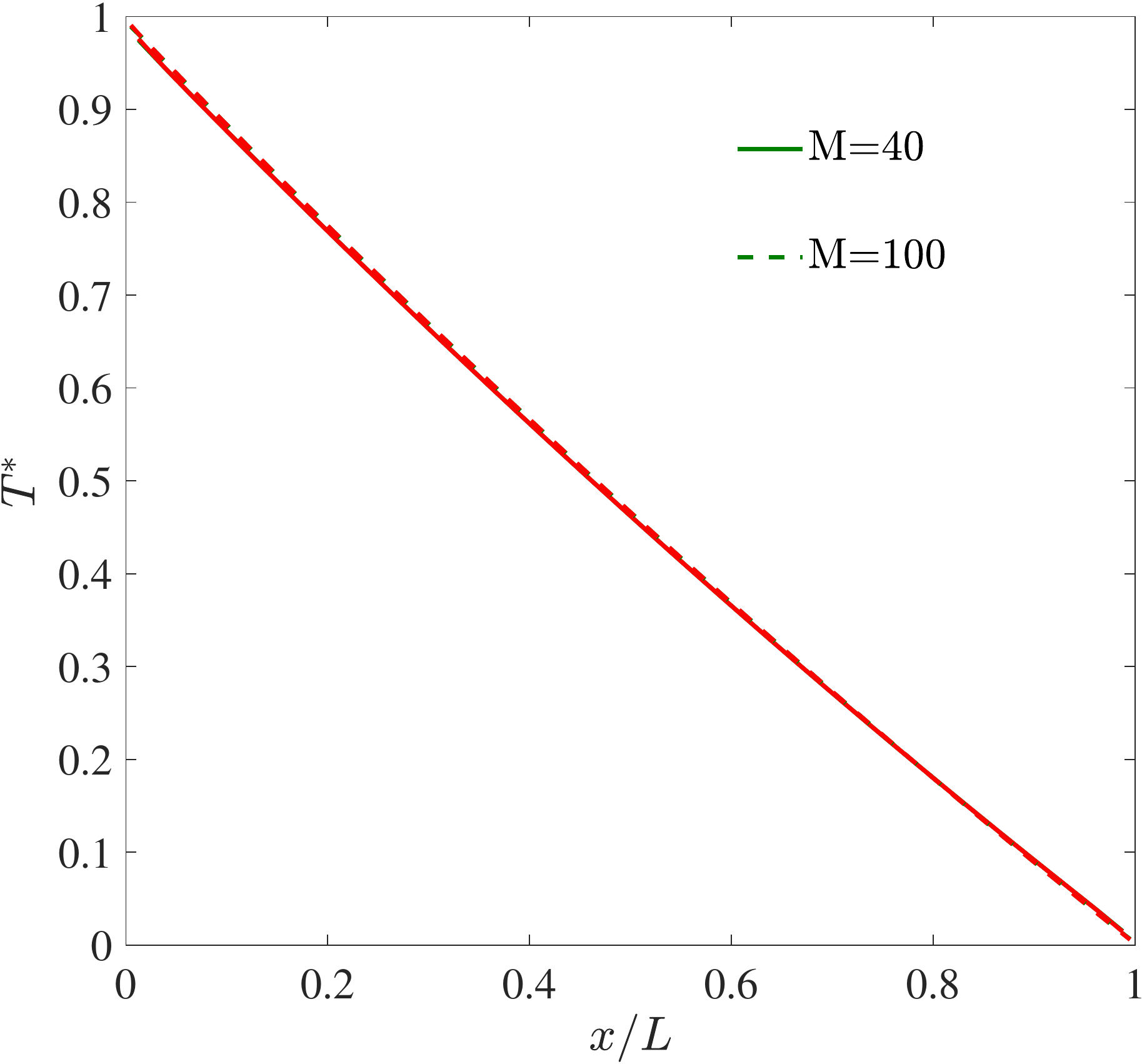} } \\
\caption{Spatial distributions of the electron and phonon temperatures with different thickness $L$, where $T^*=(T-T_R)/(T_L -T_R)$, the red line is the phonon temperature $T_p$ and the dark green is the electron temperature $T_e$. `M=40' and `M=100' represent the discretized cell number. }
\label{Au_cross_plane_M}
\end{figure}

The quasi-1D cross-plane heat conduction in Au metal film with different thickness $L$ is studied.
The wall temperature at each side of the film is fixed at $T_L= T_0 +\Delta T$ and $T_R= T_0 - \Delta T$, respectively.
The thermal physical parameters of electrons and phonons are the same as those used in~\cref{Au_film_schematic} as well as the numerical discretizations.
Spatial distributions of the electron and phonon temperatures with different thickness $L$ are shown in~\cref{Au_cross_plane}, where isothermal boundary conditions (Eq.~\eqref{eq:BC1}) are used for two walls.
It can be found that the temperature slip appears near the wall boundaries, and the temperature slip of electron is more obvious because the electron mean free path (33 nm) is larger than that of phonon (1.5 nm) in Au metals.
When the thickness increases, the electron-phonon coupling becomes frequent so that the deviations between phonon and electron temperature decreases.

We also make a test of grid independence in~\cref{Au_cross_plane_M}.
It can be found that the numerical results predicted by coarse grid (`$M=40$') are in excellent agreement with that predicted by fine grid (`$M=100$').
When the thickness $L=10~\mu$m is much larger than mean free path, the temperature slip disappears in the diffusive regime.
Furthermore, the present scheme could capture the temperature distributions accurately even the coarse cell size ($L/M=250$ nm) is much larger than the mean free path.

\subsection{Transient thermal grating}
\label{sec:TTG}

\renewcommand{\arraystretch}{1.2}
\setlength\tabcolsep{7pt}
\begin{table}[htb]
\caption{Physical parameters of BTE with coupled electron and phonon transport in TTG geometry at $300$ K and $25$ K~\cite{JHTelectron-phonon2009,PhysRevB.99.054308,Sciadv_2019_hotelectron,negative_diffusion_ACSphotonics2023}.}
\centering
\begin{tabular}{|*{3}{c|}}
 \hline
Au &  $300$ K   &  $25$ K   \\
\hline
$C_p $ (J$\cdot$m$^{-3}$$\cdot$K$^{-1}$)  &  2.50E6  &   5.10E5  \\
\hline
$\kappa_p $ (W$\cdot$m$^{-1}$$\cdot$K$^{-1}$)  &  2.75  &  23.0   \\
\hline
$ |\bm{v}_p|$ (m$\cdot$s$^{-1}$)  &  2142.857   &  2142.857  \\
 \hline
$ |\tau_p|$ (s)  &  7.187E-13    &  2.946E-11  \\
\hline
$ \lambda_p = |\bm{v}_p||\tau_p| $ (m)  & 1.540E-9   &  6.314E-8   \\
\hline
$ \alpha_p= \kappa_p /C_p $ (m$^2$$\cdot$s$^{-1}$)  & 1.10E-6  & 4.51E-5  \\
\hline
$C_e $ (J$\cdot$m$^{-3}$$\cdot$K$^{-1}$)  &  2.0E4  &  1.70E3    \\
\hline
$\kappa_e $ (W$\cdot$m$^{-1}$$\cdot$K$^{-1}$)  &  320.0   &  977.0    \\
\hline
$ |\bm{v}_e|$  (m$\cdot$s$^{-1}$)  &  1.36E6  &  1.36E6 \\
 \hline
 $ |\tau_e|$ (s)  &  2.595E-14    &  9.322E-13  \\
\hline
$ \lambda_e = |\bm{v}_e||\tau_e| $ (m)  & 3.53E-8   &  1.268E-6  \\
\hline
$ \alpha_e= \kappa_e /C_e $ (m$^2$$\cdot$s$^{-1}$)  & 0.016  &  0.575  \\
\hline
$G$ (W$\cdot$m$^{-3}$$\cdot$K$^{-1}$)   &  3.0E16  &   5.0E15   \\
\hline
\end{tabular}
\label{TTGparameters}
\end{table}
\begin{figure*}[htb]
\centering
\subfloat[$L=10$ nm]{ \includegraphics[scale=0.32,clip=true]{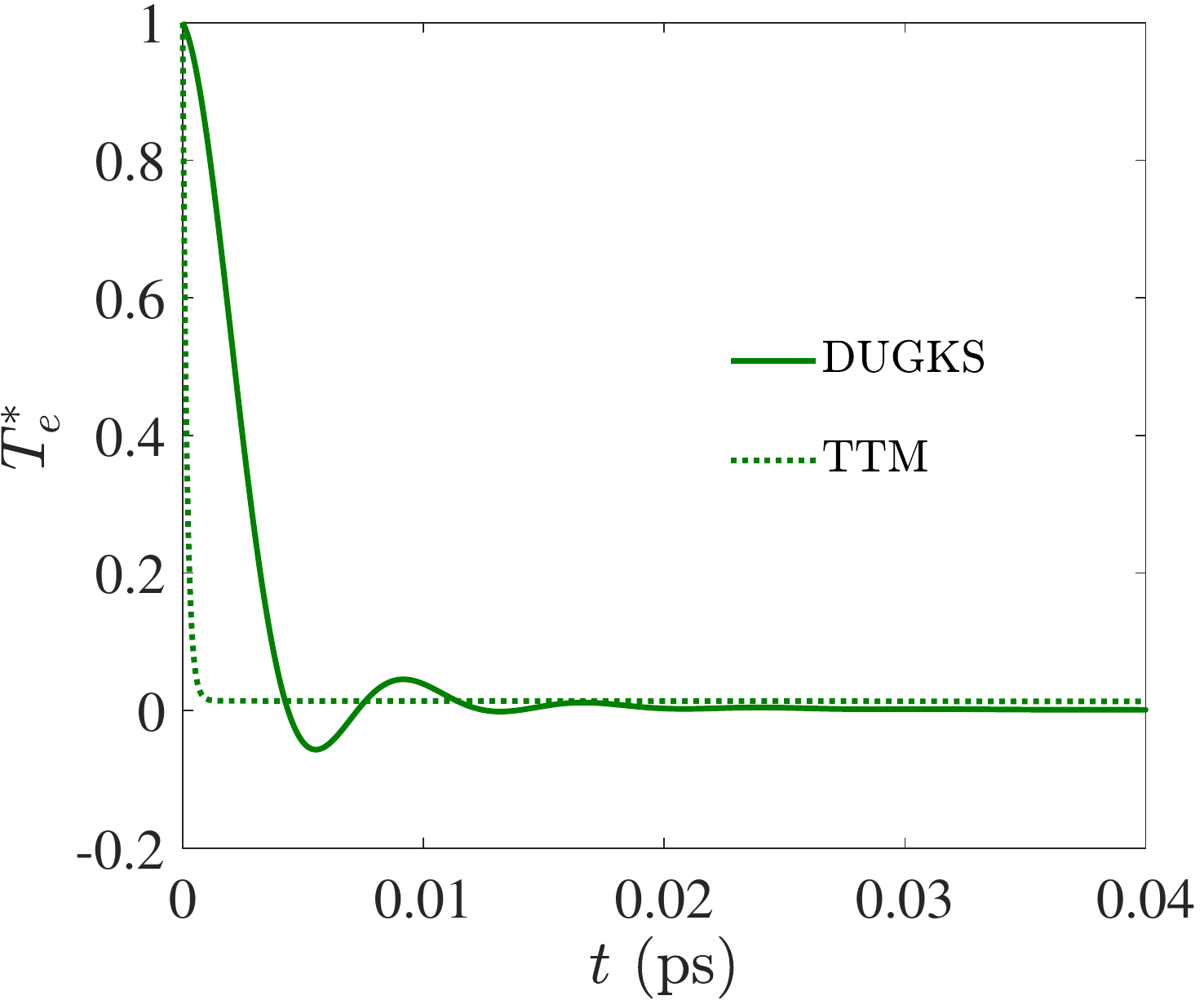} }  ~
\subfloat[$L=100$ nm]{ \includegraphics[scale=0.32,clip=true]{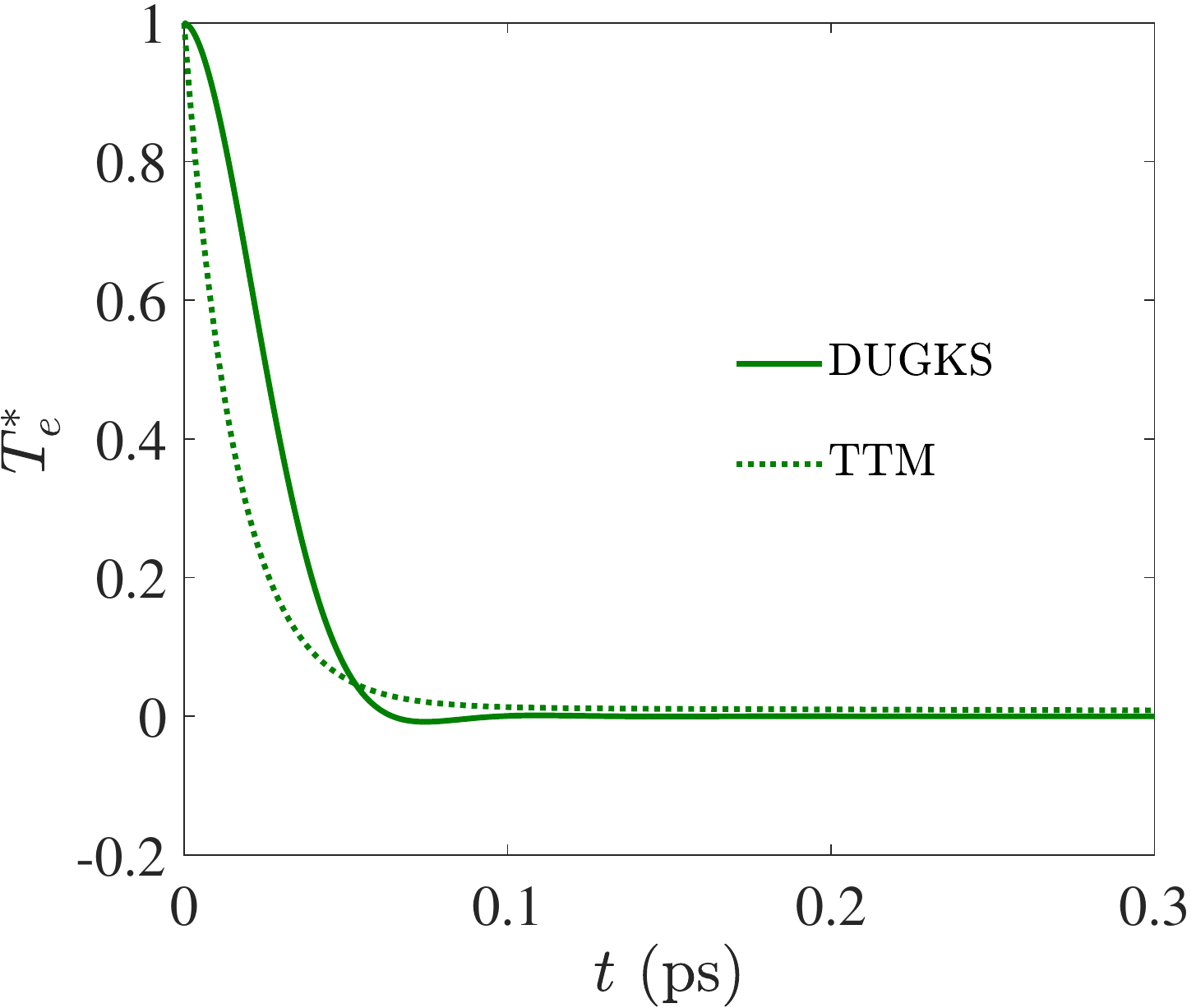} } ~
\subfloat[$L=1~\mu$m]{ \includegraphics[scale=0.32,clip=true]{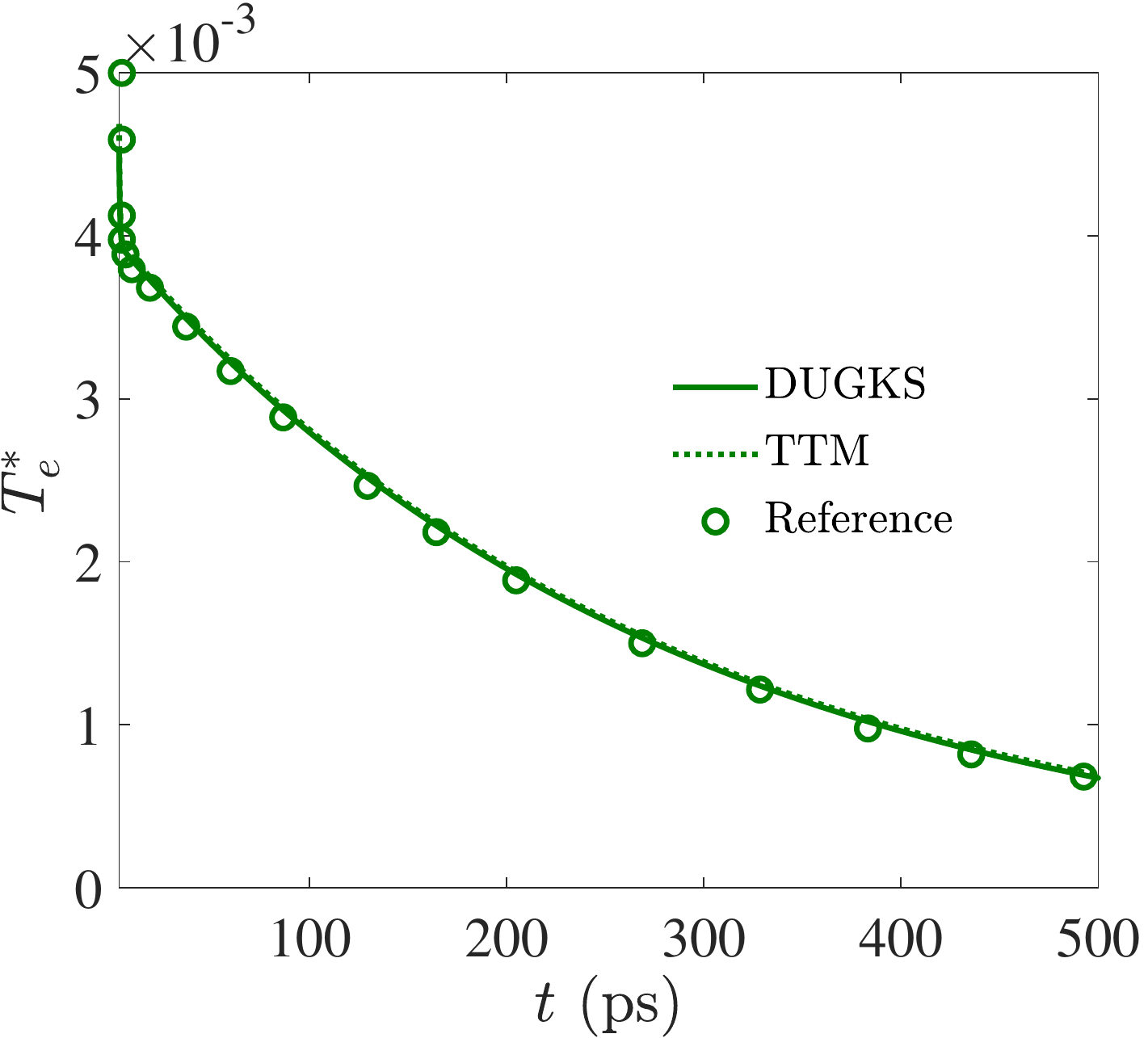} } \\
\subfloat[$L=10$ nm]{ \includegraphics[scale=0.32,clip=true]{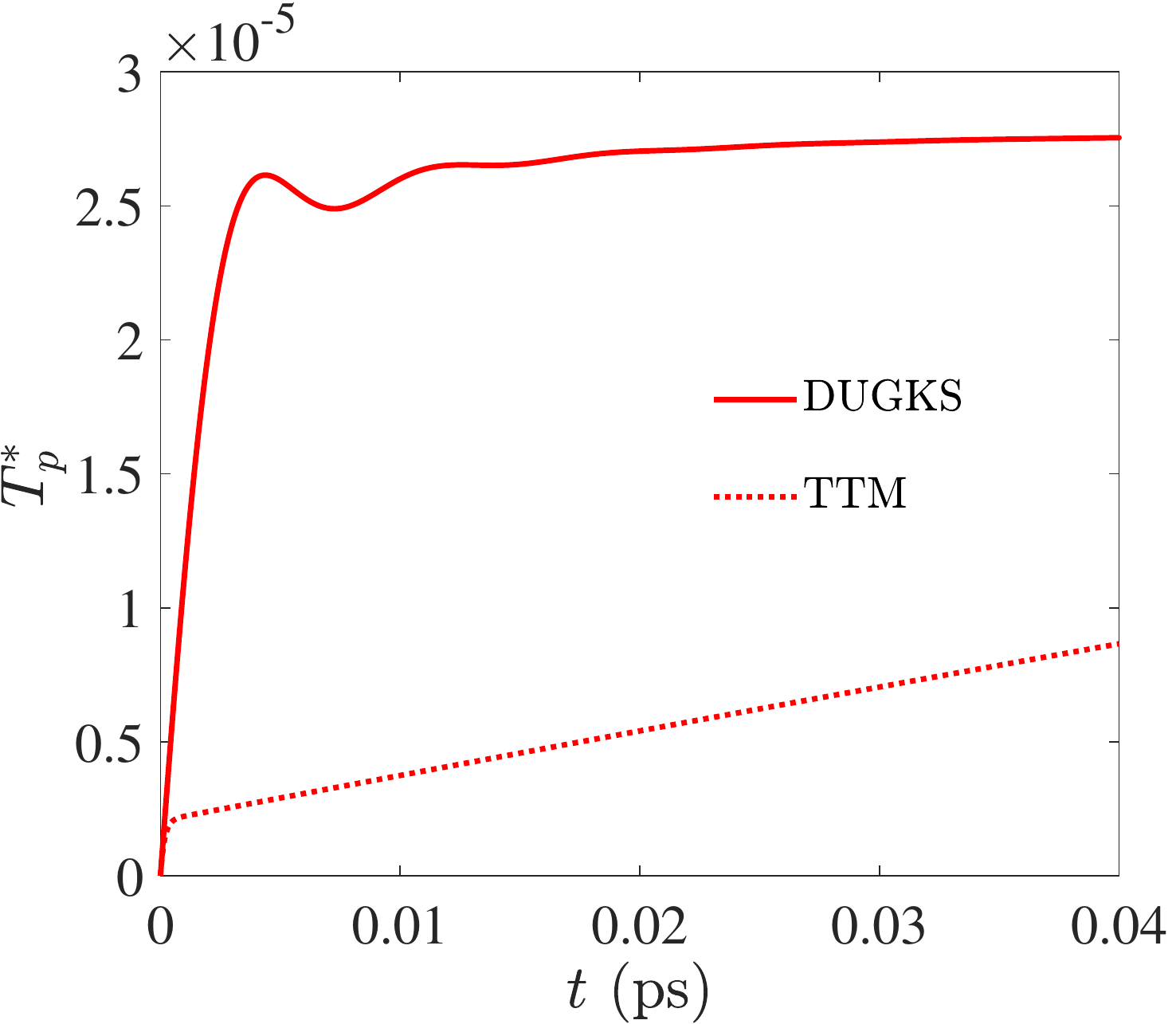} }  ~
\subfloat[$L=100$ nm]{ \includegraphics[scale=0.32,clip=true]{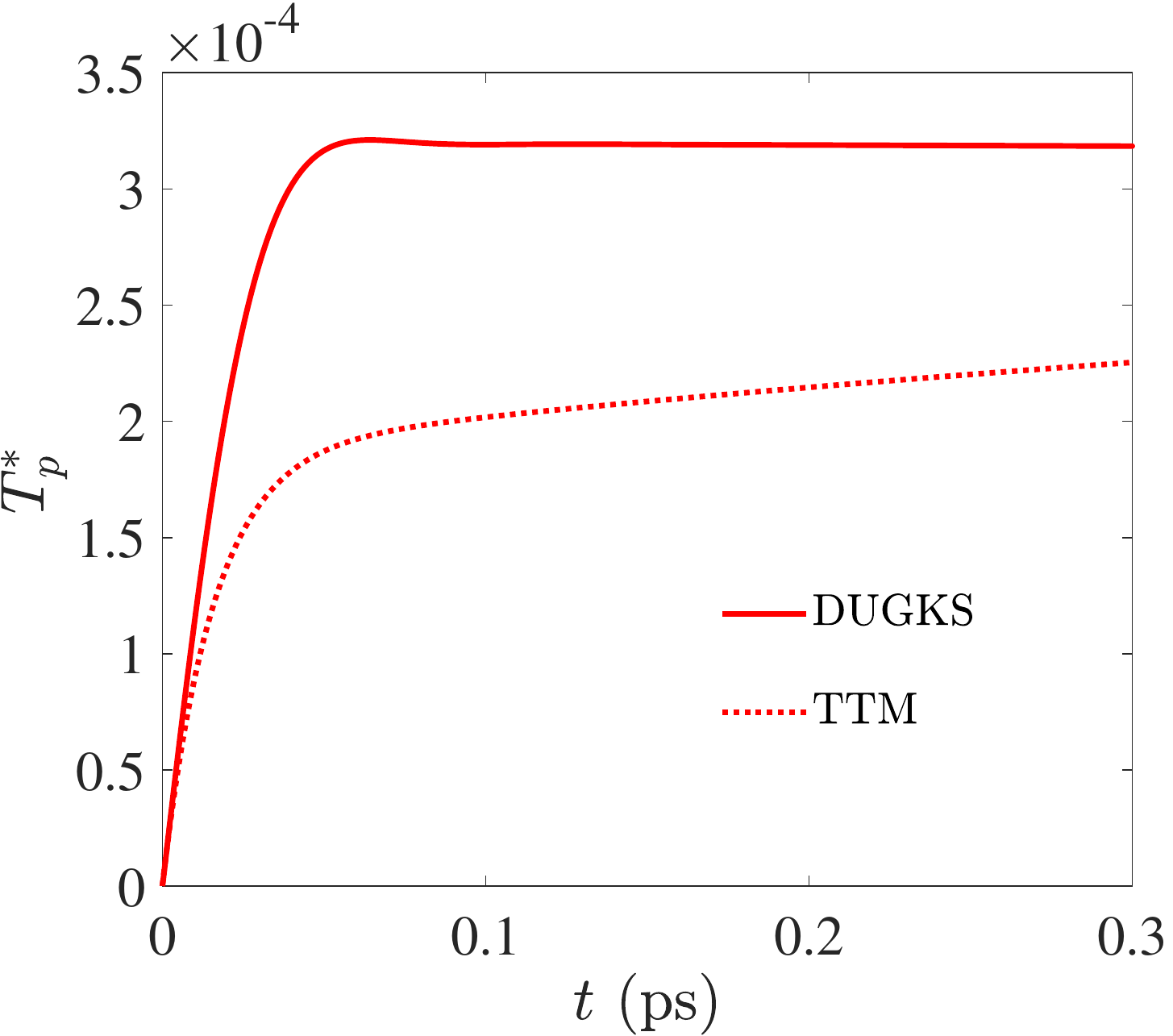} } ~
\subfloat[$L=1~\mu$m]{ \includegraphics[scale=0.32,clip=true]{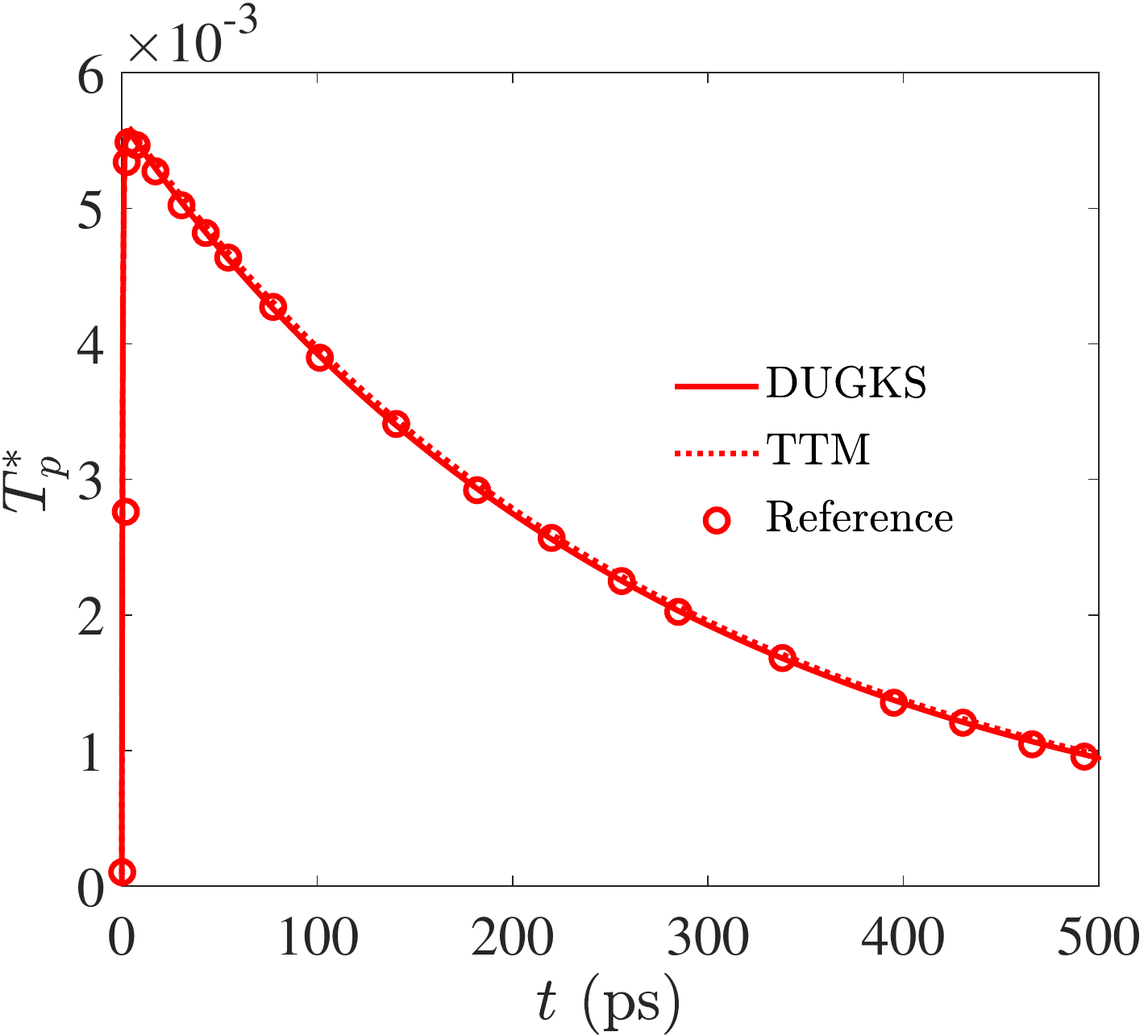} } \\
\caption{Time-dependent electron and phonon temperatures at $x=L/4$ with different $L$ when $T_{\text{ref}}=300$ K, where $\Delta T=1$ K, $T_e^*= (T_e - T_{\text{ref}})/ \Delta T$, $T_p^*= (T_p - T_{\text{ref}})/ \Delta T$. Reference data originates from Ref.~\cite{TTG_metals_2011_JAP}. }
\label{Au_TTG_300K}
\end{figure*}
\begin{figure*}[htb]
\centering
\subfloat[$L=100$ nm]{ \includegraphics[scale=0.32,clip=true]{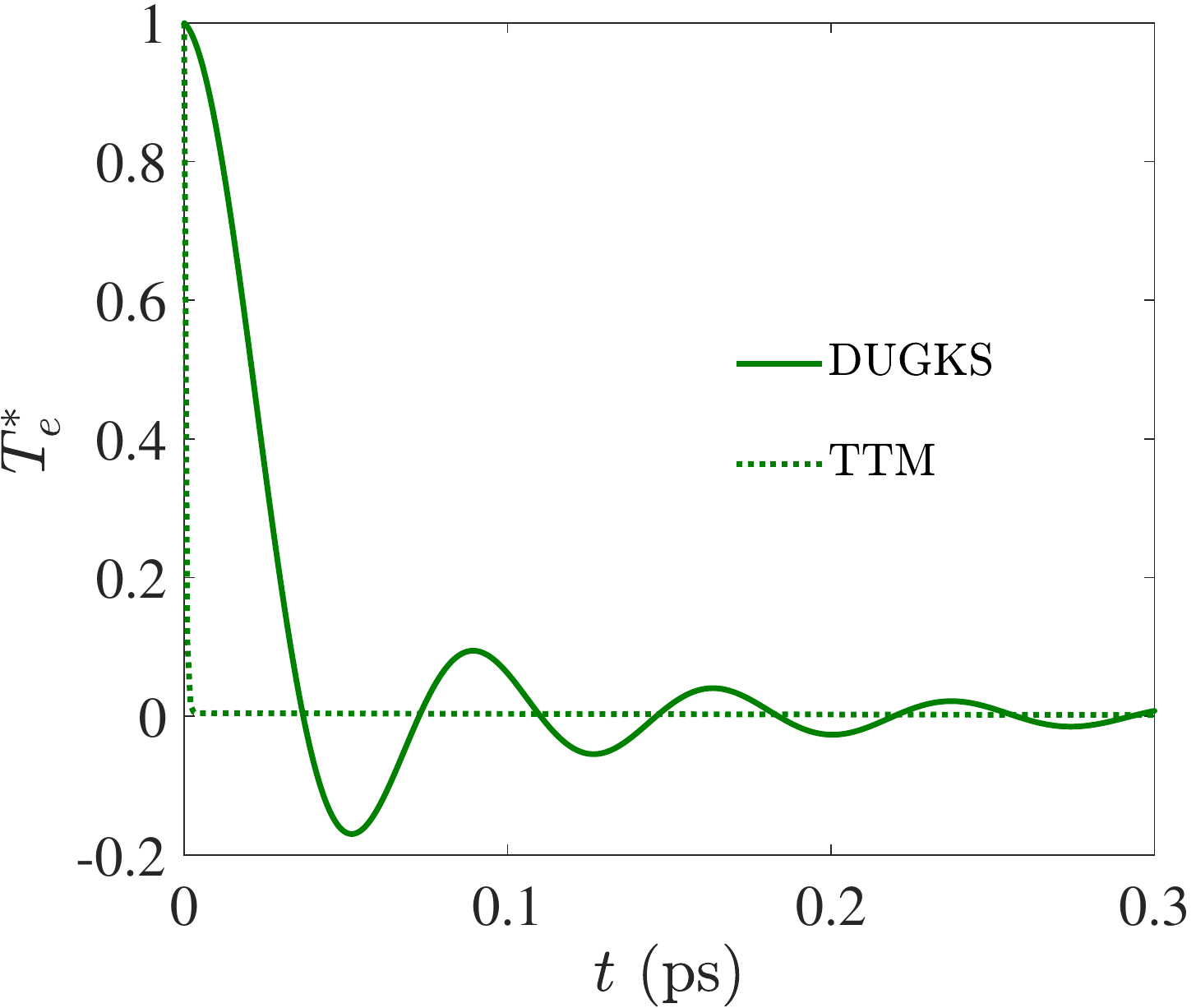} }  ~
\subfloat[$L=1~\mu$m]{ \includegraphics[scale=0.32,clip=true]{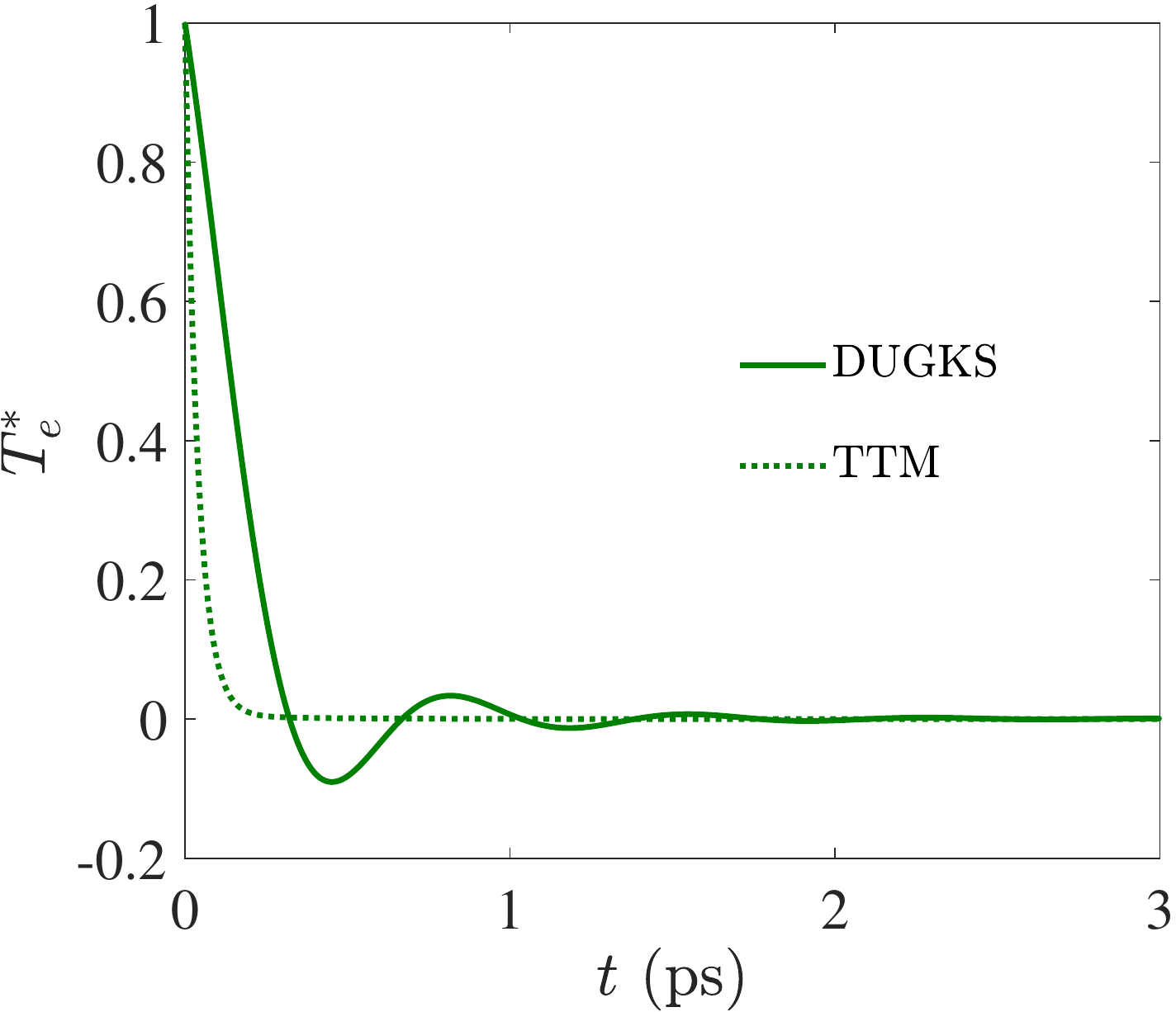} } ~
\subfloat[$L=10~\mu$m]{ \includegraphics[scale=0.32,clip=true]{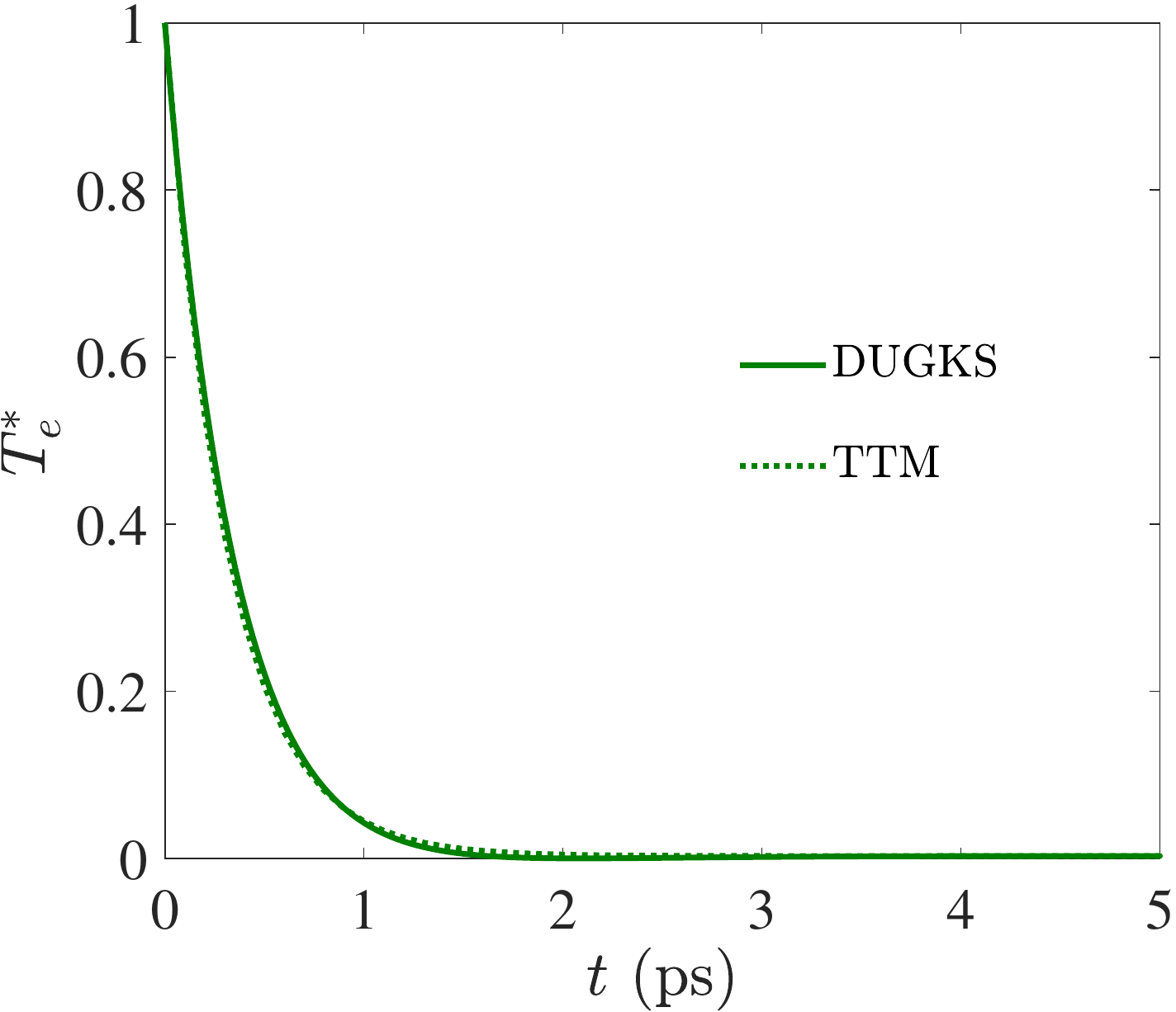} } \\
\subfloat[$L=100$ nm]{ \includegraphics[scale=0.32,clip=true]{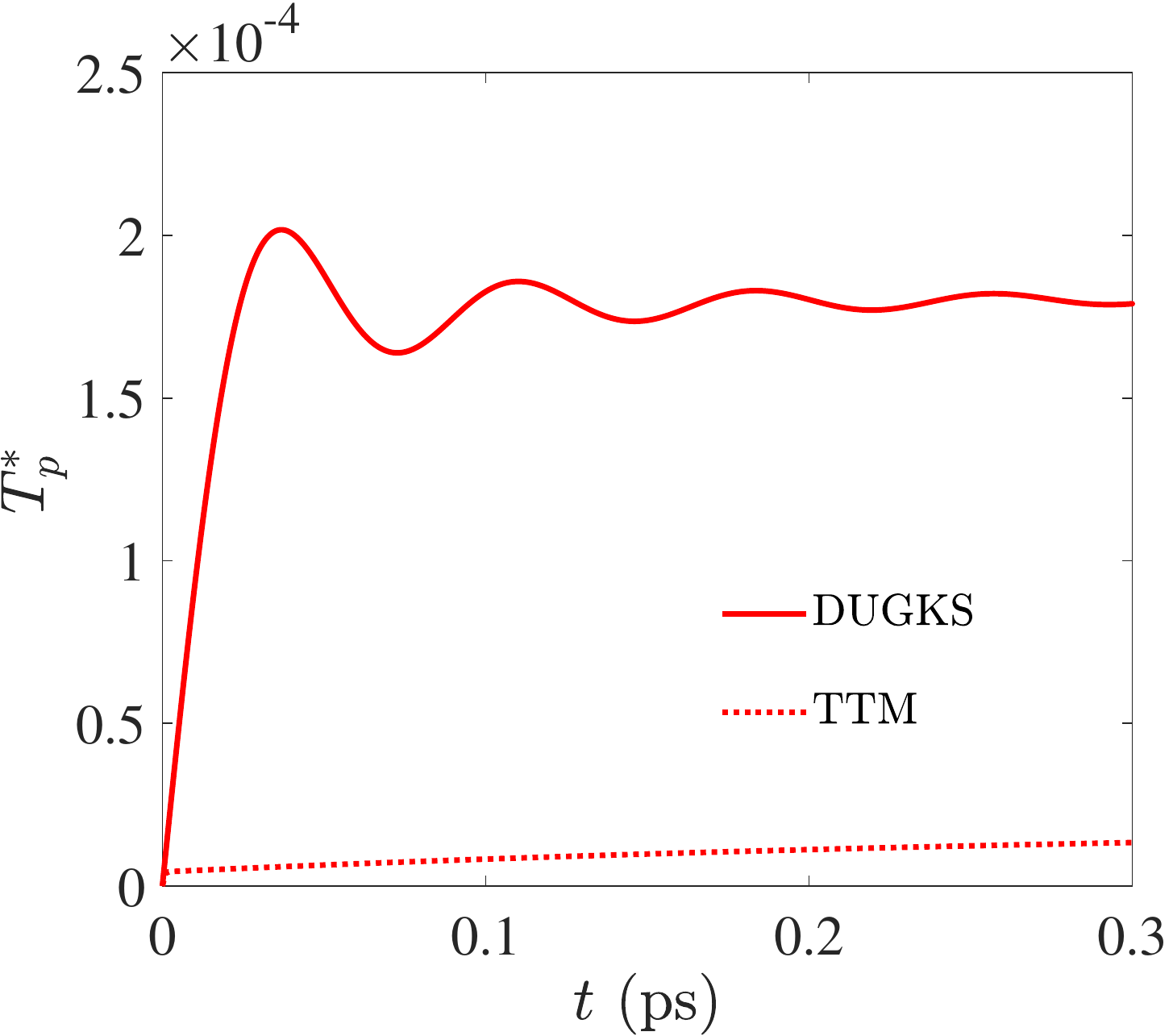} }  ~
\subfloat[$L=1~\mu$m]{ \includegraphics[scale=0.32,clip=true]{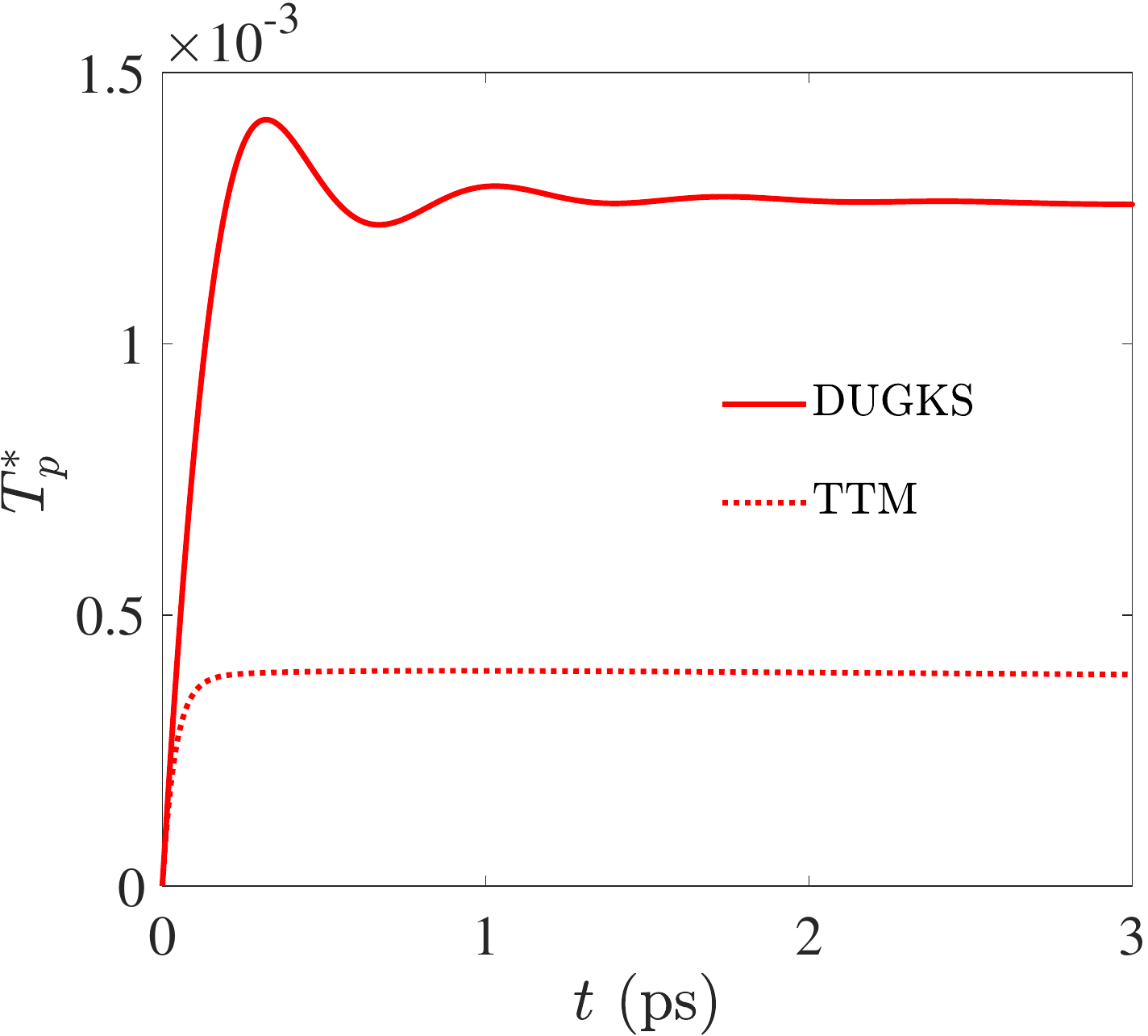} } ~
\subfloat[$L=10~\mu$m]{ \includegraphics[scale=0.32,clip=true]{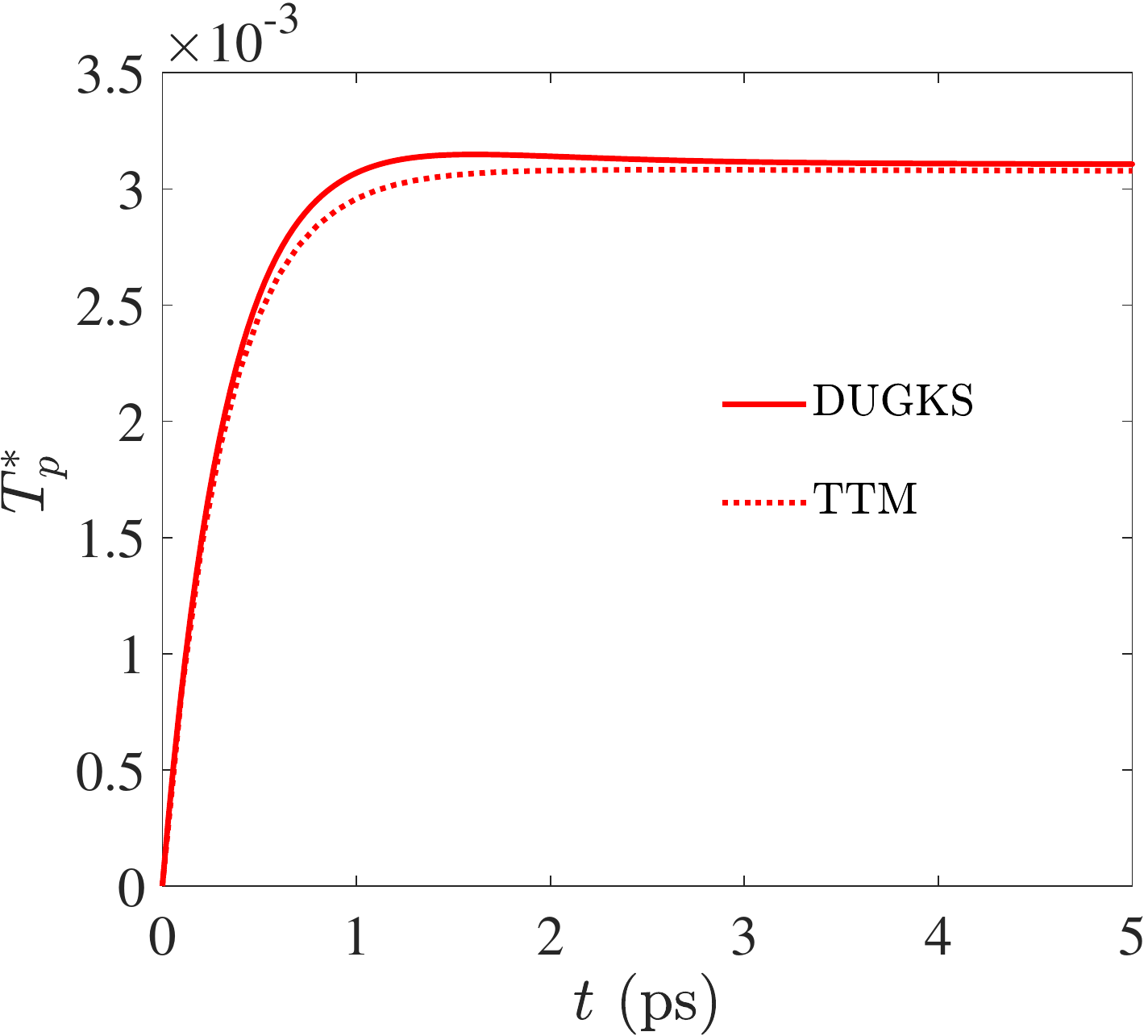} } \\
\caption{Time-dependent electron and phonon temperatures at $x=L/4$ with different $L$ when $T_{\text{ref}}=25$ K, where $\Delta T=1$ K, $T_e^*= (T_e - T_{\text{ref}})/ \Delta T$, $T_p^*= (T_p - T_{\text{ref}})/ \Delta T$.  }
\label{Au_TTG_25K}
\end{figure*}

Transient thermal grating (TTG)~\cite{TTG_metals_2011_JAP,sivan_ultrafast_2020} in a Au metal is analyzed.
In quasi-1D geometry, the crossed pump lasers produce a sinusoidal interference pattern on the surface of a sample with period $L$, which results in a spatially sinusoidal temperature profile,
\begin{align}
T_e (x,t=0) &= T_{\text{ref}} +\Delta T \sin ( 2 \pi x/L ), \\
T_p (x,t=0) &= T_{\text{ref}} ,
\end{align}
where $\Delta T$ is the temperature difference.
We assume that the grating initially heats only the electronic subsystem.
The computational domain is discretized with $20-50$ uniform cells, and $8-100$ discrete points in the $|\bm{v}| \cos \theta $ direction is used.
The periodic boundary conditions are adopted for the domain.
Detailed physical parameters of phonon and electron in Au metal are listed in Table.~\ref{TTGparameters}, where $\alpha=\kappa /C$ is the thermal diffusivity and $\lambda= |\bm{v}| \tau$ is the mean free path.

Numerical results of time-dependent electron and phonon temperatures at $x=L/4$ with different spatial grating length $L$ are shown in~\cref{Au_TTG_300K}, where $T_{\text{ref}}=300$ K and $\Delta T=1$ K.
It can be found that when $L=1~\mu$m, the DUGKS results at room temperature are the same as those obtained from Ref.~\cite{TTG_metals_2011_JAP} or predicted by the TTM, namely, the heat conduction is in the diffusive regime.
Actually the BTE could recover the TTM in the diffusive limit, see Appendix~\ref{sec:dimensionalanalysis}.
When the grating period length decreases and becomes comparable to the electron mean free path, the results predicted by the TTM deviate from the DUGKS results.
The electron temperature wave appears in the ballistic regime, which breaks the diffusion equation exactly.

Another interesting thing is that initially the electron temperature is higher than phonon temperature, so energy/heat is released from the electron to the phonon, but after a while the phonon temperature is higher than the electron temperature, namely, the energy is transferred back from the phonon to the electron, as shown in~\cref{Au_TTG_300K}(c,f).
Similar heat flow from phonon to electron $T_e - T_p <0 $ also exist when $L=10$ nm or $100$ nm based on the DUGKS results.
This phenomenon $T_e - T_p <0 $ has also been measured experimentally~\cite{negative_diffusion_ACSphotonics2023} or studied theoretically based on TTM~\cite{TTG_metals_2011_JAP} at micron scale.

The underlying physical mechanisms of $T_e - T_p <0 $ can be summarized as below.
Actually the heat conduction in electron or phonon subsystems is composed of three parts: advection $\bm{v} \cdot \nabla u$, scattering $(u^{eq} -u)/ \tau$ and electron-phonon coupling $G(T_e -T_p)$.
The purpose of advection and scattering is to make the temperature tend to be a constant at different spatial locations, i.e., $T(\bm{x}) \rightarrow  T_{\text{ref}}$.
While the electron-phonon coupling aims to establish a local equilibrium between phonon and electron subsystems for a given spatial position, i.e., $ |T_e -T_p|  \rightarrow  0$.
In the diffusive regime, the phonon-phonon and electron-electron scattering are sufficient so that the heat dissipations depend on the thermal diffusivity $\alpha$ and electron-phonon coupling $G$.
In the ballistic regime, the phonon-phonon and electron-electron scattering are rare so that the heat dissipations depend on the phonon/electron advection speed $|\bm{v}|$ and electron-phonon coupling $G$.
The thermal diffusivity and group velocity of electron are much larger than those of phonon, which indicates that the phonon subsystem is more like a energy accumulator relative to the electron subsystem in both diffusive and ballistic regimes.
Hence the electron temperature varies between phonon temperature $T_p$ and reference temperature $T_{\text{ref}}$.

When the reference temperature decreases from $300$ K to $25$ K, both thermal diffusivity and mean free path increases.
When $L \leq 1~\mu$m, electron ballistic transport is obvious and temperature wave appears.
The TTM results deviate significantly from the DUGKS results, because the diffusive transport is no longer valid due to the small grating period size.
It can also be found that the anomalous thermal phenomena $T_e - T_p <0 $ exist.
When $L=10~\mu$m, the TTM results are in consistent with the DUGKS data, and the thermal transport is in the diffusive regime.

Although the TTM is valid in the diffusive regime, the numerical results in~\cref{Au_TTG_25K}(c,f) are quite different from those in~\cref{Au_TTG_300K}(c,f).
In~\cref{Au_TTG_300K}(c,f), the electron-phonon coupling dominates heat dissipation at the initial stage so that the electron temperature decreases significantly, and then the electron and phonon temperatures decay slowly due to the small thermal diffusivity.
In~\cref{Au_TTG_25K}(c,f), the thermal diffusivity of electron and phonon are much larger than that at room temperature, so that it can be found that both electron and phonon temperature tend to a constant in a shorter time.

\subsection{Thermal transport in a metallic bilayer}
\label{sec:bilayer}

\begin{figure}[htb]
\centering
\includegraphics[scale=0.55,clip=true]{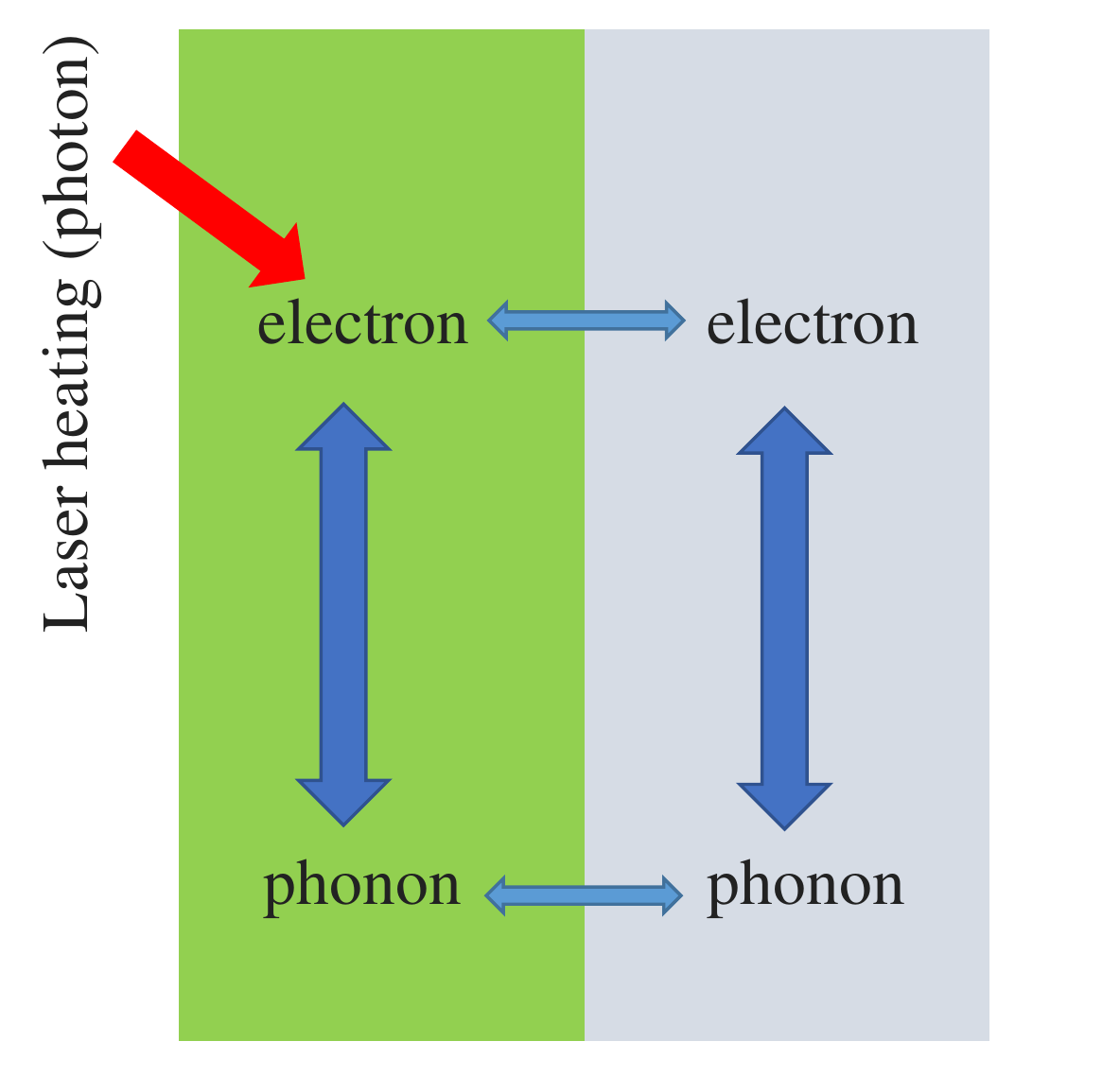}
\caption{Schematic of the physical evolution process of electron/phonon transport in bilayer metals with interfacial thermal resistance. }
\label{bilayer_schematic}
\end{figure}
\begin{table}[htb]
\caption{Thermal physical parameters of electrons and phonons in Pt metals~\cite{JAP2016EP_review} at $300$ K.}
\centering
\begin{tabular}{|*{2}{c|}}
\hline
$C_p $ (J$\cdot$m$^{-3}$$\cdot$K$^{-1}$)  &  2.67E6     \\
\hline
$\kappa_p $ (W$\cdot$m$^{-1}$$\cdot$K$^{-1}$)  &  5.80   \\
\hline
$ |\bm{v}_p|$ (m$\cdot$s$^{-1}$)  &  1.91E3  \\
\hline
$C_e $ (J$\cdot$m$^{-3}$$\cdot$K$^{-1}$)  &   $\gamma T_e$   \\
\hline
$\gamma$ J$\cdot$m$^{-3}$$\cdot$K$^{-2}$  &  $748.1$     \\
\hline
$\kappa_e $ (W$\cdot$m$^{-1}$$\cdot$K$^{-1}$)  & $ \kappa_0 T_e /T_p $  \\
\hline
$\kappa_0 $ (W$\cdot$m$^{-1}$$\cdot$K$^{-1}$)  & $ 65.80 $  \\
\hline
$ |\bm{v}_e|$  (m$\cdot$s$^{-1}$)  &  0.46E6    \\
\hline
$G$ (W$\cdot$m$^{-3}$$\cdot$K$^{-1}$)   &  108.80E16  \\
\hline
\end{tabular}
\label{Pt_parameters}
\end{table}
\begin{figure}[htb]
\centering
\subfloat[]{ \includegraphics[scale=0.28,clip=true]{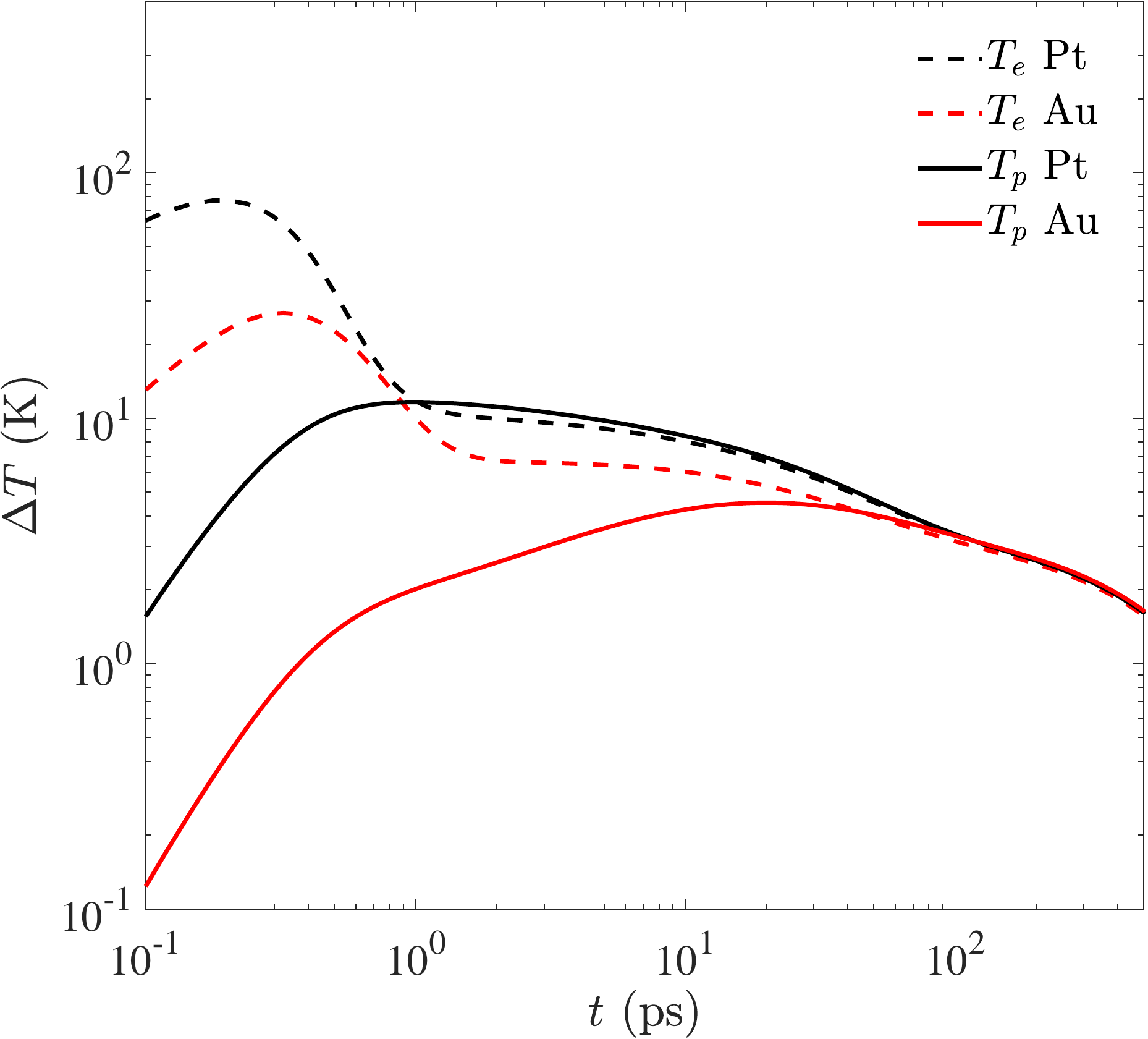} }  \\
\subfloat[]{ \includegraphics[scale=0.28,clip=true]{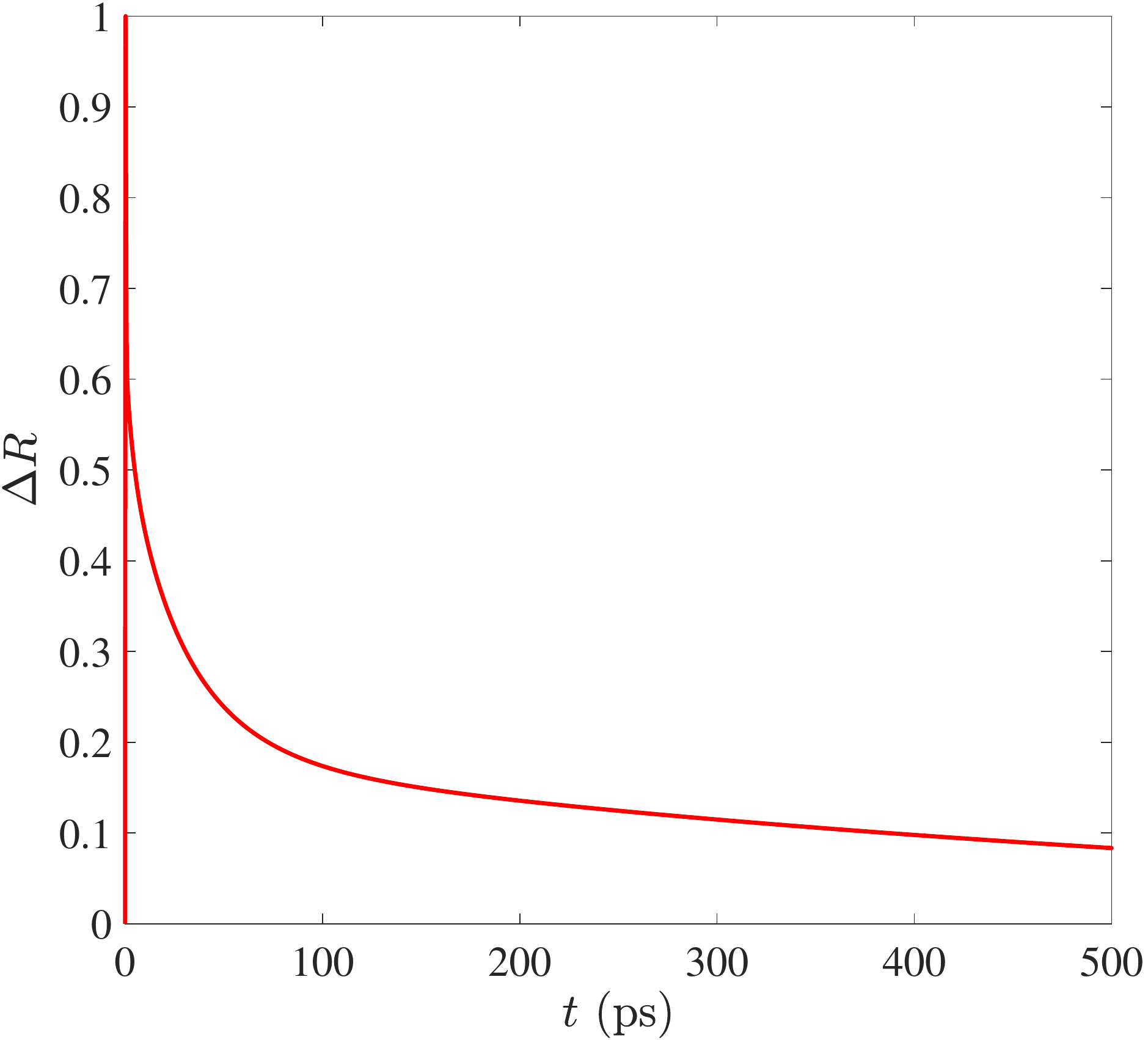} }  \\
\caption{Transient (a) temperature evolution and (b) reflected signal in Pt/Au bilayer metals, where $\Delta T= T-T_{\text{ref}} $ and $\Delta R= a \Delta T_e + b \Delta T_p$. For Pt, $a/b=0.25$.  }
\label{BilayerPtAu}
\end{figure}
\begin{figure}[htb]
\centering
\subfloat[Au/Pt]{ \includegraphics[scale=0.28,clip=true]{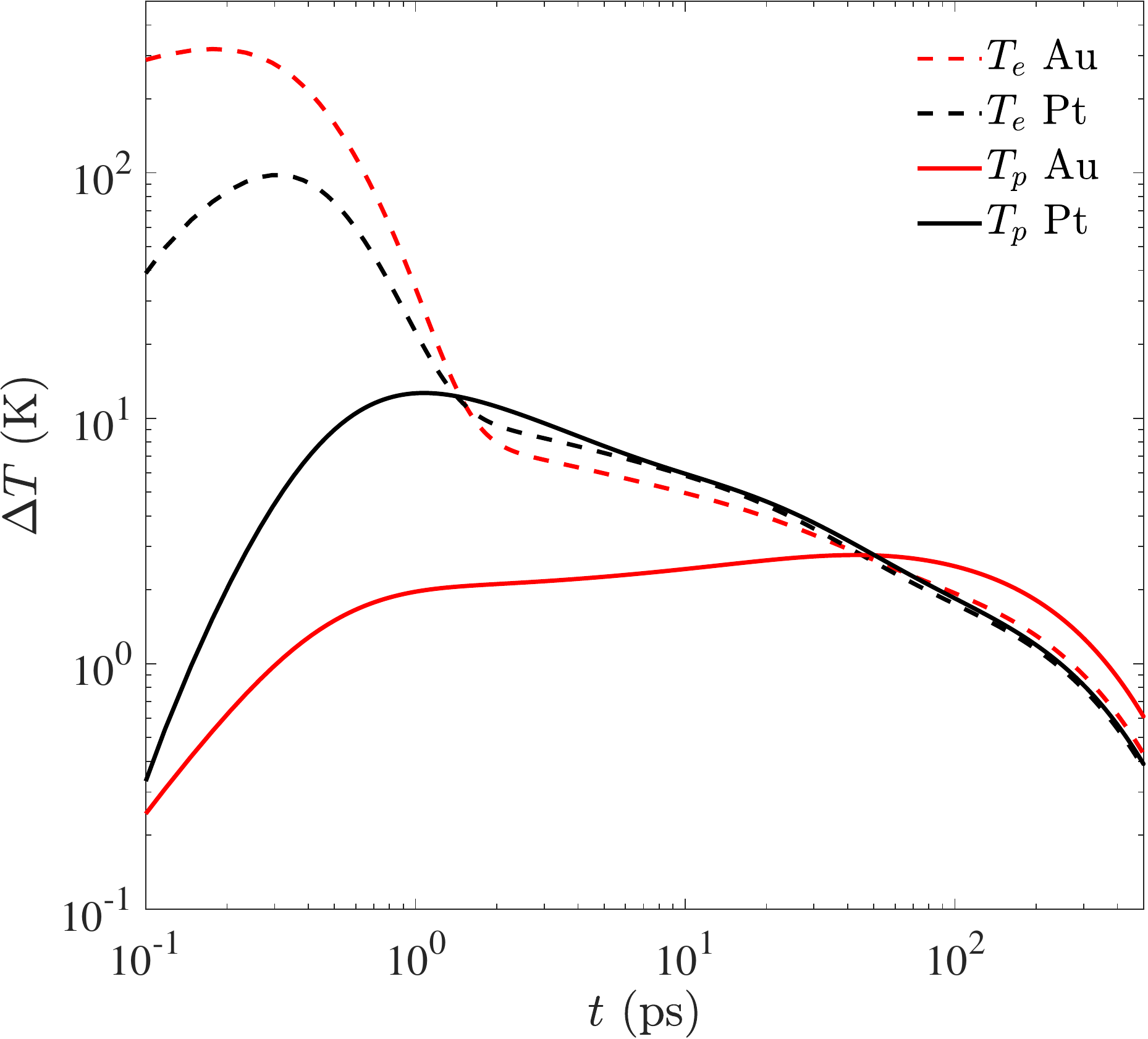} } \\
\subfloat[Au/Pt]{ \includegraphics[scale=0.28,clip=true]{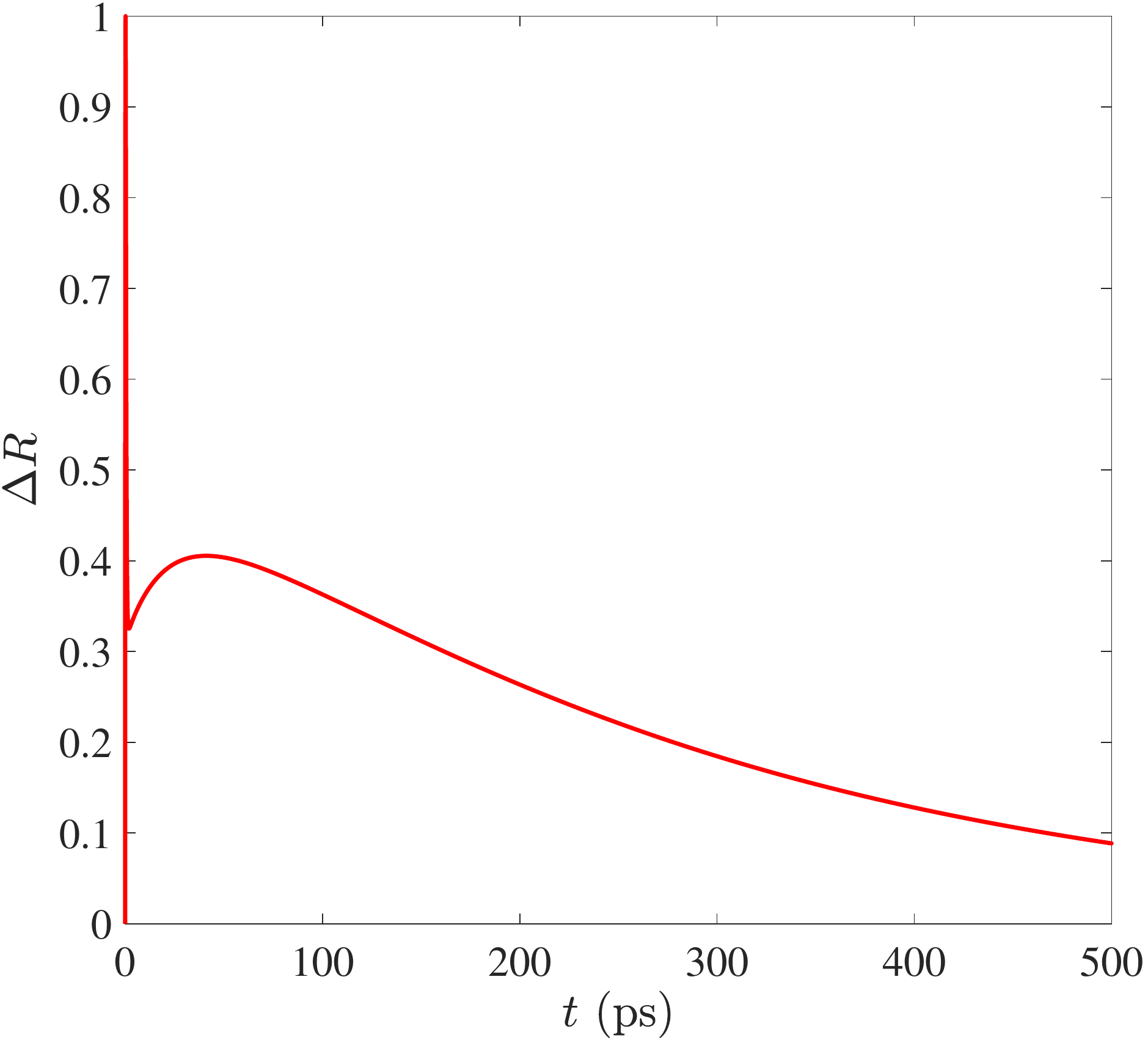} }   \\
\caption{Transient (a) temperature evolution and (b) reflected signal in Au/Pt bilayer metals, where $\Delta T= T-T_{\text{ref}} $ and $\Delta R= a \Delta T_e + b \Delta T_p$. For Au, $a/b=0.02$.  }
\label{BilayerAuPt}
\end{figure}
\begin{figure*}[htb]
\centering
\subfloat[$t=20\Delta t$]{ \includegraphics[scale=0.28,clip=true]{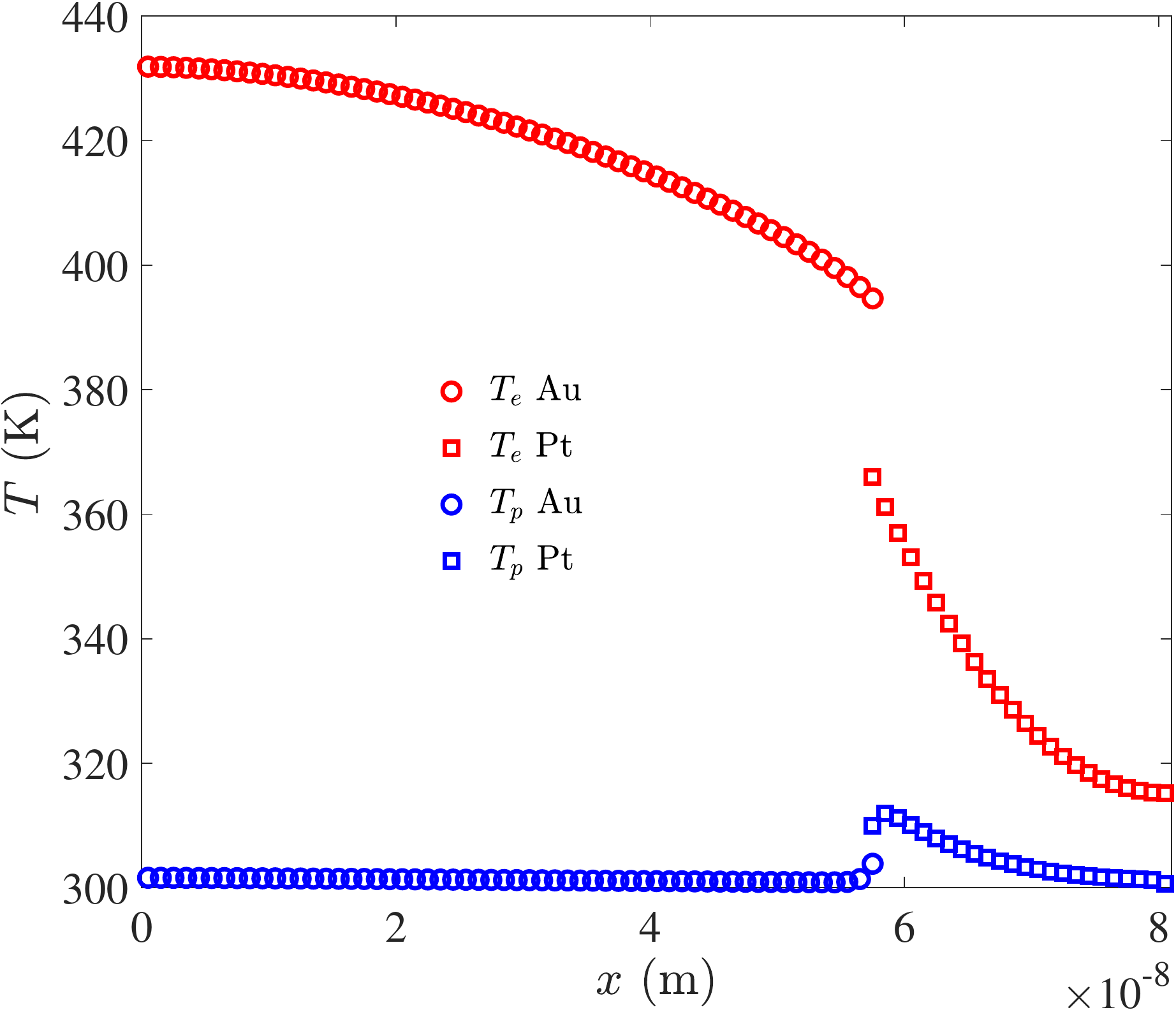} }  ~
\subfloat[$t=60\Delta t$]{ \includegraphics[scale=0.28,clip=true]{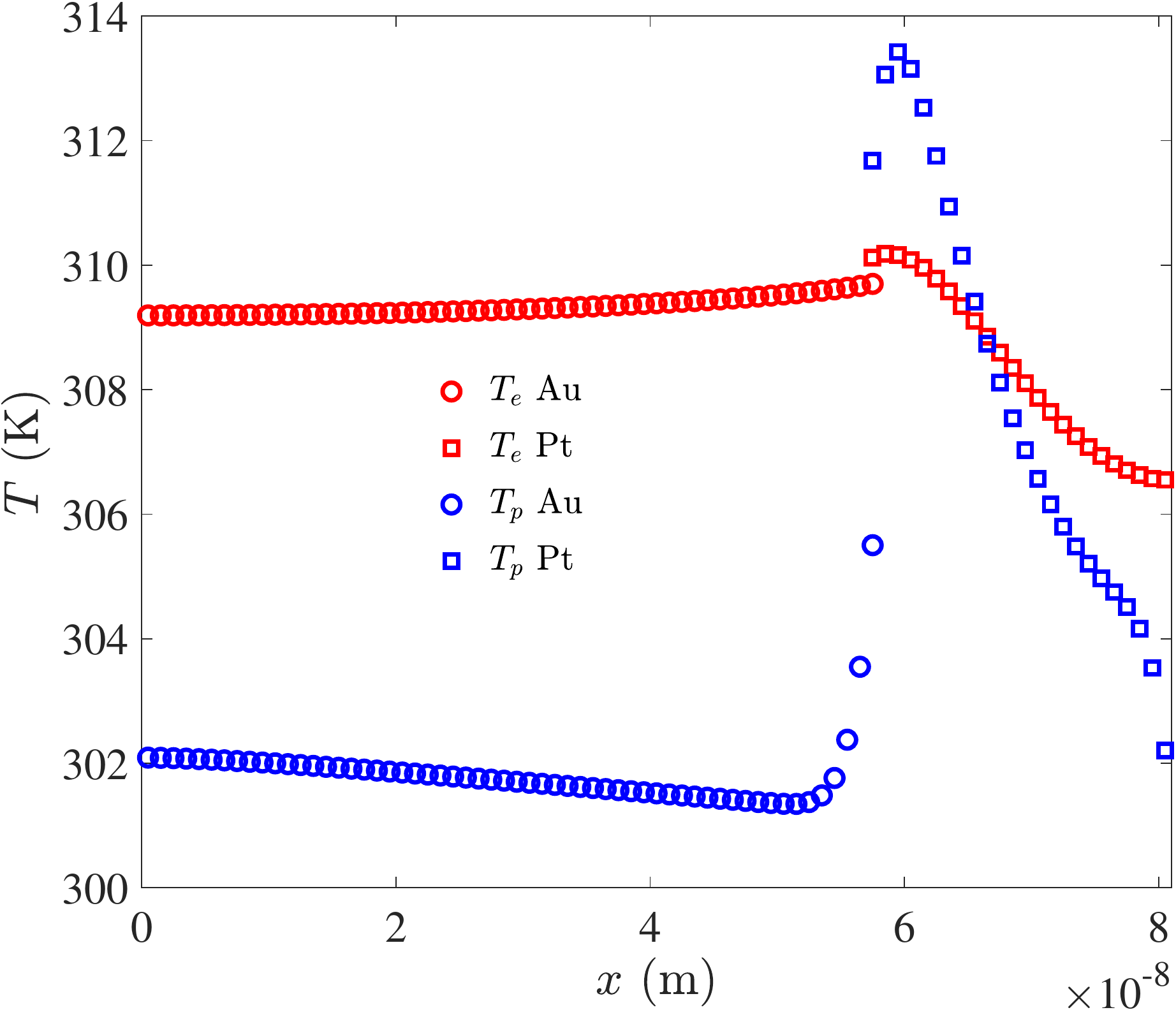} }  \\
\subfloat[$t=1000\Delta t$]{ \includegraphics[scale=0.28,clip=true]{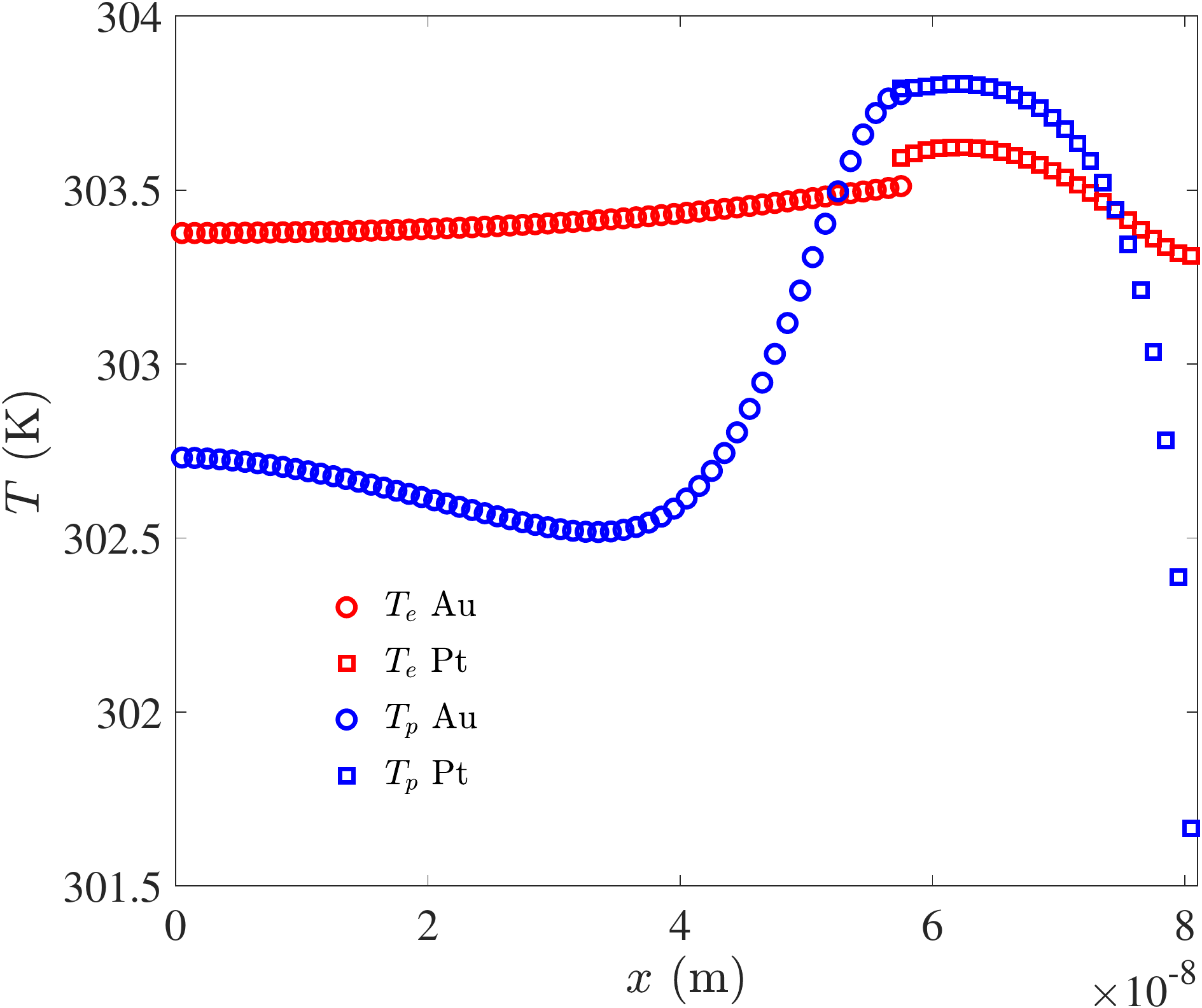} }  ~
\subfloat[$t=10000\Delta t$]{ \includegraphics[scale=0.28,clip=true]{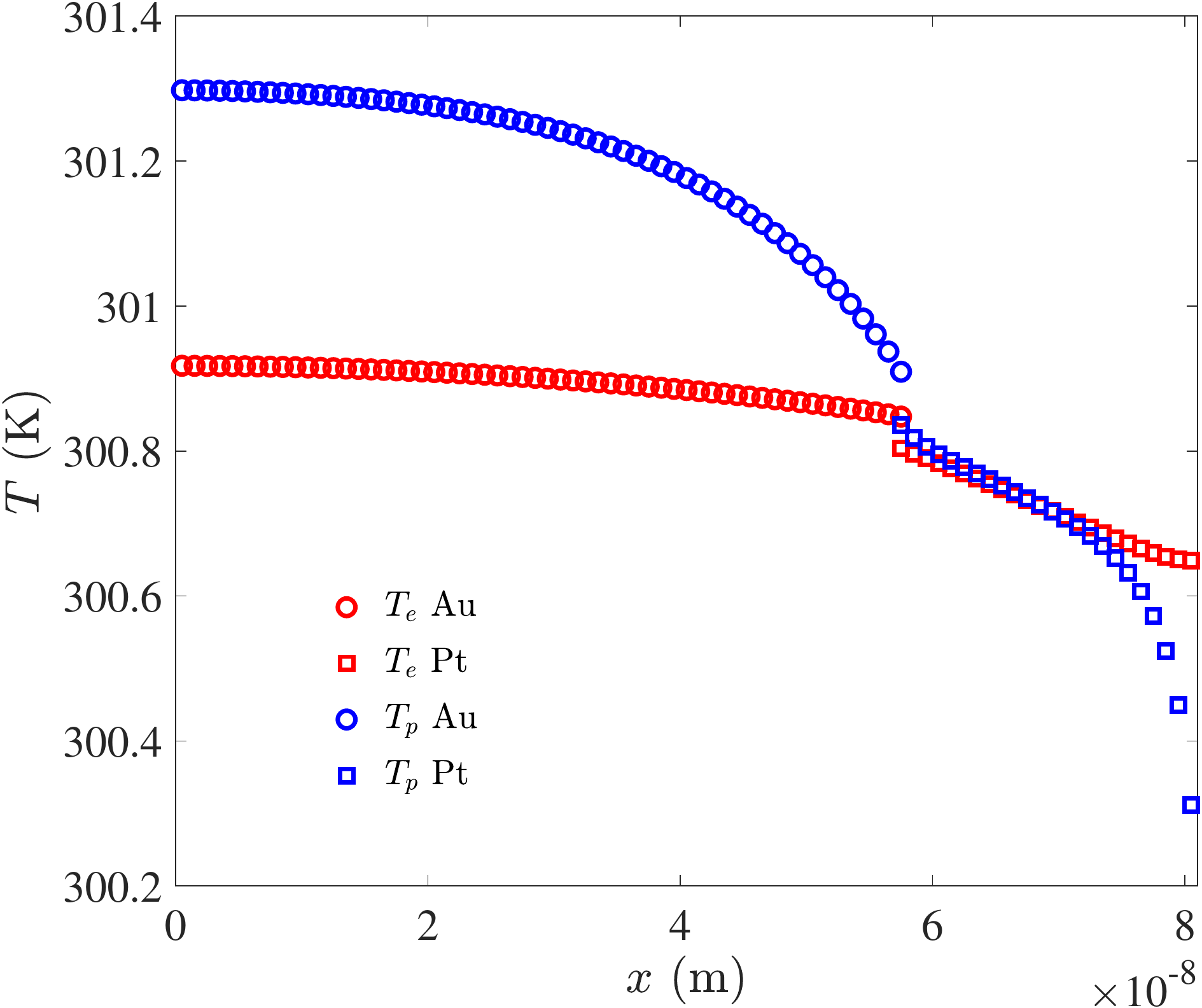} }  \\
\caption{Spatial distributions of electron and phonon temperature in Au/Pt at different moment, where $\Delta t= 0.0294 $ ps. }
\label{Bilayer_tempperature}
\end{figure*}

The heat conduction in the last four subsections only consider a single-layer metal film.
In the practical experiments or applications, the heat conduction in bilayer or multilayer dissimilar materials is more common~\cite{JAP2015EPC,hopkins_JAP_2009,karna_direct_2023,PhysRevB.89.064307}.
In this subsection, the thermal transport in a metallic bilayer (Au/Pt or Pt/Au~\cite{PhysRevB.89.064307}) with ultrafast laser heating is studied by the DUGKS and the interfacial thermal resistance between two dissimilar materials are accounted~\cite{RevModPhys.94.025002,JAP2016EP_review,MIAO2023123538}.

The diffuse mismatch model~\cite{RevModPhys.94.025002,JAP2016EP_review,MIAO2023123538}, in which electron/phonon scattering at the metal-metal interface is assumed to be completely diffuse, is used to describe the interfacial thermal resistance.
The transmittance $t$ and reflectance $r$ on each side of the interface satisfy the following constrictions due to energy conservation,
\begin{align}
r_{ij} + t_{ij} &=1,
\end{align}
where $t_{ij}$ represents the transmittance from medium $i$ to medium $j$ across the interface, and $r_{ij}$ represents the reflectance in the medium $i$ reflected back from the interface.
In addition, at the thermal equilibrium state (the temperature of the left and right side of interface is the same), the net heat flux across the interface is zero due to the principle of detailed balance so that
\begin{align}
t_{ij} C_i v_i  =  t_{ji} C_j v_j,
\end{align}
where $C_i$ and $v_i$ are the specific heat and group velocity of medium $i$.
We only consider the interaction between electron (phonon) in medium $i$ and electron (phonon) in medium $j$.
The schematic of particle transport in bilayer metals is shown in~\cref{bilayer_schematic}.

We consider a combination of a 23 nm Pt film and a 58 nm Au film with initial environment temperature $T_{\text{ref}}= 300$ K.
The thermal physical parameters of electrons and phonons in Au metals are shown in previous Tab.~\ref{TTGparameters} with $\gamma =71.0$ J$\cdot$m$^{-3}$$\cdot$K$^{-2}$, and the thermal physical parameters of electrons and phonons in Pt metals are shown in Tab.~\ref{Pt_parameters}~\cite{JAP2016EP_review}.

Firstly, we simulate the transient heat conduction in Au/Pt bilayer metal, and the external heat source is shown in Eq.~\eqref{eq:sourcepump} with $t_{pump}= 400$ fs and $P_{input} (1-R_r ) = 1.60 $ J$\cdot$m$^{-2}$~\cite{PhysRevB.89.064307}.
The specular boundary conditions are adopted for the heating top surface regardless phonon and electron.
For the unheated bottom surface, the specular boundary condition (Eq.~\eqref{eq:BC2}) is used for electron transport and the isothermal boundary condition (Eq.~\eqref{eq:BC1}) with environment temperature $300$ K is used for phonon transport, which indicates that the net electron heat flux at the bottom surface is zero and the phonon heat flux is nonzero.
The uniform cell is used with cell size $1$ nm, and the CFL number is $0.80$.
$40$ discrete points in the $|\bm{v}| \cos \theta $ direction is used.
Similarly, we also simulate the transient heat conduction in Pt/Au bilayer metal.

Numerical results are shown in~\cref{BilayerPtAu}.
It can be found that in Pt/Au metals, the phonon temperature in Pt is always higher than the phonon temperature in Au.
The reflected signal increases firstly and then decreases gradually with time.
For Au/Pt metals the thermal behaviors are completely different as shown in~\cref{BilayerAuPt}.
It can be found that the phonon temperature in the unheated metal is higher than the phonon temperature in the heated metal in the early tens of picoseconds.
The reflected signal is also non-monotonic with time after the laser heating is almost removed.
It increases with time in the early tens of picoseconds and finally decreases with time.

In order to understand this anomalous heat conduction phenomenon, the spatial distributions of electron and phonon temperature in Au/Pt bilayer metal at different moment are plotted in~\cref{Bilayer_tempperature}.
At the early stage $t=20\Delta t$, the electron temperature in Au is highest.
There are obvious temperature slip or interfacial thermal resistance in the Au/Pt interface for both electron and phonon transport.
In addition, the phonon temperature in the Pt side is higher than that in the Au side.
That's because the electron-phonon coupling of Pt is much stronger than that of Au.
When $t=60\Delta t$, the electron temperature in Pt side near the Au/Pt interface is higher than the electron temperature in Au side.
When $t=1000\Delta t$, the phonon temperature in Au metal increases and the lattice energy come from two part.
One part is the electron-phonon coupling in Au because the electron temperature is still higher than the phonon temperature.
The other is the phonon heat flows back from the Pt to Au side, which leads to that the deviations of the phonon temperature on the each side of Au/Pt interface decreases.
When $t=10000\Delta t$, the electron-phonon, electron-electron, phonon-phonon scattering are sufficient so that all temperatures decreases with time gradually as shown in~\cref{BilayerAuPt}(a).

\section{Conclusion}
\label{sec:conclusion}

Non-equilibrium thermal conduction in ultrafast heating systems is studied by the BTE accounting for electron-phonon coupling.
A discrete unified gas kinetic scheme is developed to directly solve the BTE, in which the electron/phonon advection, scattering and electron-phonon interactions are coupled together within one time step by solving the BTE again at the cell interface.
Numerical results show that the present scheme not only correctly predicts the heat conduction in the diffusive regime with coarse cell size, but also captures the ballistic or thermal wave effects when the characteristic length is comparable to or smaller than the mean free path where the TTM fails.
For ultrafast laser heating problem, the present results are in excellent agreement with the experimental results in existing literatures and our performed TDTR.

In transient thermal grating geometry, heat flow from phonon to electron is predicted in both the ballistic and diffusive regimes.
It results from the competition of the thermal diffusivity and electron-phonon coupling in the diffusive regime, and in the ballistic regime it results from the competition of the phonon/electron advection and electron-phonon coupling.
Furthermore, in Au/Pt bilayer metals with interfacial thermal resistance, the predicted reflected signal increases in the early tens of picoseconds and then decreases with time after the heat source is removed.
The stronger electron-phonon coupling strength in the unheated Pt side results in the higher phonon temperature in Pt, so that phonon heat flows back from Pt to Au side.

\section*{Acknowledgments}

This work is supported by the China Postdoctoral Science Foundation (2021M701565).
The authors are grateful to Prof. Yonatan Sivan in Ben-Gurion University and Dr. Wuli Miao in Tsinghua University for communications on electron-phonon coupling.


\onecolumngrid

\appendix

\section{Two-temperature model}
\label{sec:TTM}

Two-temperature model (TTM)~\cite{TTM_review2010,QIU19942789,PhysRevB.77.075133,PhysRevB.65.214303} with coupled electron and phonon thermal transport is
\begin{align}
C_e \frac{ \partial T_e }{\partial t  } &= \nabla \cdot ( \kappa_e \nabla T_e )   -G(T_e -T_p)  + S  ,\\
C_p \frac{ \partial T_p }{\partial t  } &= \nabla \cdot ( \kappa_p \nabla T_p )   +G(T_e -T_p)  .
\end{align}
In the discretized time space, the diffusion term and heat source term are dealt with the semi-implicit scheme and the electron-phonon coupling term is implemented with explicit scheme,
\begin{align}
&C_e^m \frac{ \delta T_e^{m} }{\Delta t}-0.5\nabla \cdot ( \kappa_e^{m} \nabla \delta T_e^{m} ) \notag \\
&= \nabla \cdot ( \kappa_e^{m} \nabla T_e^{m} ) - G(T_e^{m} -T_p^{m} ) +  \frac{S^m+S^{m+1} }{2}  ,\\
&C_p^m \frac{ \delta T_p^{m} }{\Delta t  }-0.5\nabla \cdot ( \kappa_p^{m} \nabla  \delta T_p^{m} ) \notag  \\
&= \nabla \cdot ( \kappa_p^{m} \nabla T_p^{m} )  + G(T_e^{m} -T_p^{m} ) ,
\end{align}
where $\delta T^{m} =T^{m+1}- T^{m}$ is the temperature increment for each time step $m$.
Actually the electron-phonon coupling term can also be implemented with semi-implicit scheme,
\begin{align}
&C_e^m \frac{ \delta T_e^{m} }{\Delta t}-0.5\nabla \cdot ( \kappa_e^{m} \nabla \delta T_e^{m} ) + 0.5G(\delta T_e^{m} -\delta  T_p^{m} ) \notag \\
&= \nabla \cdot ( \kappa_e^{m} \nabla T_e^{m} ) - G(T_e^{m} -T_p^{m} ) +  \frac{S^m+S^{m+1} }{2}  ,\\
&C_p^m \frac{ \delta T_p^{m} }{\Delta t  }-0.5\nabla \cdot ( \kappa_p^{m} \nabla  \delta T_p^{m} )- 0.5G(\delta T_e^{m} -\delta T_p^{m} ) \notag  \\
&= \nabla \cdot ( \kappa_p^{m} \nabla T_p^{m} )  + G(T_e^{m} -T_p^{m} ) .
\end{align}
But the latter method will generate a larger coefficient matrix with coupled electron and phonon temperature increment $G(\delta T_e^{m} -\delta  T_p^{m} )$ compared to the former.
We adopt the first strategy in this work for the simplicity.
To simulate the present quasi-1D numerical cases, $10-100$ uniform cells are used and the central scheme is used to discrete the diffusion term in the spatial space.

\section{Boundary conditions}
\label{sec:bcBTE}

Boundary conditions are one of the key parts in the numerical simulations.
Two kinds of boundary conditions are used in this work.
\begin{enumerate}
    \item
    Thermalization/isothermal boundary assumes that the incident particles (phonons/electron) are all absorbed by the boundary $\bm{x}_{b}$, and the particles emitted from the boundary are the equilibrium state with the boundary temperature $T_{b}$, i.e.,
    \begin{equation}
    u(\bm{x}_{b} )=u^{eq}(T_{b}), \quad \bm{v} \cdot \mathbf{n}_{b} > 0,
    \label{eq:BC1}
    \end{equation}
    where $\mathbf{n}_{b}$ is the normal unit vector of the boundary pointing to the computational domain.
    \item
    The specular reflecting boundary condition is
    \begin{align}
    u(\bm{v} )= u(\bm{v}' ),  \quad \bm{v}' \cdot \mathbf{n}_{b} <0,
    \label{eq:BC2}
    \end{align}
    where $\bm{v}' = \bm{v} - 2 \mathbf{n}_{b} (\bm{v} \cdot \mathbf{n}_{b} )$ is the incident direction.
\end{enumerate}

\section{Temperature dependent thermoreflectance signals}
\label{sec:reflectedsignal}

The reflectivity signal $R$~\cite{Sciadv_2019_hotelectron,Nonlinear_Thermore2011JHT} for a thin Au film with thickness $h$ and refractive index ($n_2 =\sqrt{\epsilon}$) under perpendicular incidence from air ($n_1 =1 $) and supported on a sapphire substrate ($n_3 =1.7$) is $R=|r|^2$, where
\begin{align}
r &= \frac{r_{12} + r_{23}  \cdot \exp( 2 i \beta ) }{1+ r_{12}   r_{23}  \cdot \exp( 2 i \beta ) },  \\
\beta  &= 2 \pi n_2 h /\lambda_0 ,  \\
r_{jk} &= \frac{ n_j  - n_k}{n_j  - n_k } ,
\end{align}
where $i$ is the imaginary number.
$\lambda_0$ is the wave length of probe pulse.
The permittivity or dielectric function $\epsilon$ of Au thin film is related to the electron temperature $T_e$ and phonon temperature $T_p$, i.e.,
\begin{align}
\epsilon &=\epsilon_{\infty}  -\frac{ \omega_p^2 (T_p) }{ \omega_0 ( \omega_0 + i \gamma_{re}(T_e,T_p) )  },  \\
\gamma_{re}(T_e,T_p) &= A_{ee} T_e^2 +B_{ep}  T_p ,
\end{align}
where $\omega_p \approx 1.37 \times 10^{16}$ rad/s is the plasma frequency~\cite{kittel1996introduction,Liang_JHT2012}, $\epsilon_{\infty} =9.50$,
$A_{ee} = 1.77 \times 10^7 $ K$^{-2}$ $ \cdot$s$^{-1}$,  $B_{ep}=1.45 \times 10^{11} $ K$^{-1}$ $ \cdot$s$^{-1}$.
$\omega_0$ is the angular frequency of probe pulse.
The nonlinear thermoreflectance model is complicated, hence in many references the change in the reflected signal is usually assumed to be linear with the increase in temperature.

\section{Dimensional analysis}
\label{sec:dimensionalanalysis}

Make a dimensional analysis of the BTE (\ref{eq:epBTE1},\ref{eq:epBTE2}) without external heat source,
\begin{align}
\frac{ \partial u_e^* }{\partial t^*} + |\bm{v}_e^*| \bm{s} \cdot \nabla_{\bm{x}^*} u_e^* &= \frac{ u_e^{eq,*}  -u_e^* }{\tau_e^* } -  \frac{ (T_e^* -T_p^* )/(4 \pi) }{1/G^*  } ,  \label{eq:dimensionalessbte1}   \\
\frac{ \partial u_p^* }{\partial t^*} + |\bm{v}_p^*| \bm{s}  \cdot \nabla_{\bm{x}^*} u_p^* &= \frac{ u_p^{eq,*}  -u_p^* }{\tau_p^* } +  \frac{ (T_e^* -T_p^* )/(4 \pi) }{1/G^*  },  \label{eq:dimensionalessbte2}
\end{align}
where $\bm{s}$ is the unit directional vector and the thermal physical parameters of electrons are regarded as the reference variables $v_{\text{ref} } =|\bm{v}_e|$,  $C_{\text{ref} }= C_e$, so that
\begin{align}
t^* &= \frac{t}{t_{\text{ref} }},  \quad& G^* &= \frac{ t_{\text{ref} }  }{ C_{\text{ref} }/G   } , \quad&  \bm{x}^* &= \frac{ \bm{x}  }{ L_{\text{ref} }},   \\
u_p^*&= \frac{u_p}{ C_{\text{ref} } T_{\text{ref} }  },  \quad&  \tau_p^* &= \frac{ \tau_p  }{ t_{\text{ref} }}, \quad&  \bm{v}_p^* &= \frac{ \bm{v}_p  }{ v_{\text{ref} }},  \\
u_e^*&= \frac{u_e}{ C_{\text{ref} } T_{\text{ref} }  },  \quad&  \tau_e^* &= \frac{ \tau_e  }{ t_{\text{ref} }}, \quad&  \bm{v}_e^* &= \frac{ \bm{v}_e  }{ v_{\text{ref} }},  \\
T_p^* &= \frac{T_p}{T_{\text{ref}} } ,  \quad&  T_e^* &=\frac{T_e}{T_{\text{ref} }}, \quad &  t_{\text{ref} }&= \frac{ L_{\text{ref} } }{v_{\text{ref}} },
\end{align}
where $L_{\text{ref} }$ is the system characteristic size.
When $t^* \gg \tau_e^*$ and $x^* \gg  |\bm{v}_e^*| \tau_e^*$, diffusive electron transport happens.
When $t^* \gg \tau_p^*$ and $x^* \gg  |\bm{v}_p^*| \tau_p^*$, diffusive phonon transport happens.
When $t^* \gg 1/G^*$, $x^* \gg |\bm{v}_e^*|/G^*$ and $x^* \gg |\bm{v}_p^*|/G^*$, electron-phonon coupling is sufficient so that the deviations between electron and phonon temperatures tend to zero.

When both electron and phonon suffer diffusive transport processes, according to the first-order Chapman-Enskpg expansion the distribution function can be approximated as
\begin{align}
u_e^*  &\approx  u_e^{eq,*} - \tau_e^* \left( \frac{ \partial u_e^{eq,*} }{\partial t^*} + |\bm{v}_e^*| \bm{s} \cdot \nabla_{\bm{x}^*} u_e^{eq,*}  +\frac{ (T_e^* -T_p^* )/(4 \pi) }{1/G^*  }   \right) , \label{eq:btece1}  \\
u_p^*  &\approx  u_p^{eq,*} - \tau_p^* \left( \frac{ \partial u_p^{eq,*} }{\partial t^*} + |\bm{v}_p^*| \bm{s} \cdot \nabla_{\bm{x}^*} u_p^{eq,*}  - \frac{ (T_e^* -T_p^* )/(4 \pi) }{1/G^*  }   \right) . \label{eq:btece2}
\end{align}
Combined above six equations and taking an integral of the BTE (\ref{eq:dimensionalessbte1},\ref{eq:dimensionalessbte2}) over the whole first Brillouin zone, we can get
\begin{align}
\frac{ \partial  }{\partial t^* } \left< u_e^{eq,*} - \tau_e^* \left( \frac{ \partial u_e^{eq,*} }{\partial t^*}  +\frac{ (T_e^* -T_p^* )/(4 \pi) }{1/G^*  }   \right)   \right> + \nabla_{\bm{x}^*} \cdot \left< |\bm{v}_e^*||\bm{v}_e^*| \bm{s} \bm{s} \tau_e^* u_e^{eq,*} \right> & =-  \frac{  T_e^* -T_p^*   }{1/G^*  }  , \\
\frac{ \partial  }{\partial t^* } \left< u_p^{eq,*} - \tau_p^* \left( \frac{ \partial u_p^{eq,*} }{\partial t^*}  +\frac{ (T_e^* -T_p^* )/(4 \pi) }{1/G^*  }   \right)   \right> + \nabla_{\bm{x}^*} \cdot \left< |\bm{v}_p^*||\bm{v}_p^*| \bm{s} \bm{s} \tau_p^* u_p^{eq,*} \right> & =  \frac{  T_e^* -T_p^*   }{1/G^*  }  ,
\end{align}
where $<>$ represents the integral over the whole first Brillouin zone.
Then the associated macroscopic heat conduction is
\begin{align}
C_e \frac{ \partial T_e  }{\partial t }  - \tau_e  \frac{ \partial^2 U_e  }{\partial t^2 } - \tau_e \frac{ \partial (G T_e - G T_p) }{\partial t }     &= \nabla \cdot ( \kappa_e \nabla T_e )  -  G(T_e -T_p)  , \label{eq:TTMce1}    \\
C_p \frac{ \partial T_p  }{\partial t }  - \tau_p  \frac{ \partial^2 U_p  }{\partial t^2 } + \tau_p \frac{ \partial (G T_e - G T_p) }{\partial t }     &= \nabla \cdot ( \kappa_e \nabla T_p )  +  G(T_e -T_p) , \label{eq:TTMce2}
\end{align}
where $\kappa = \int C |\bm{v}|^2 \tau /3  d\bm{K}$ is the bulk thermal conductivity.
When $\sqrt{\tau_e^*} \ll t^*$, $\sqrt{\tau_p^*} \ll t^*$, $\tau_e^* /(t^*/G^*) \ll 1$, $\tau_p^* /(t^*/G^*) \ll 1$, the first order term of relaxation time in Eqs.~(\ref{eq:TTMce1},\ref{eq:TTMce2}) could be removed and above two equations recover the typical two-temperature model in the diffusive limit.

\bibliography{phonon}

\end{document}